\documentclass[acmsmall,screen]{acmart}

\AtBeginDocument{%
  }

\usepackage{xspace}
\usepackage{graphicx}
\usepackage{algorithm}
\usepackage{algpseudocode}
\usepackage[capitalise]{cleveref}
\usepackage[table]{xcolor}
\usepackage{pifont}
\usepackage{tabularx}
\usepackage{booktabs}
\usepackage{multirow}
\usepackage{array}
\usepackage{fancybox}
\usepackage{enumitem}
\usepackage{makecell}

\usepackage{fvextra}
\DefineVerbatimEnvironment{MintedVerbatim}{Verbatim}{commandchars=\\\{\}}
\input{_minted/default.style.minted}

\newcolumntype{C}{>{\centering\arraybackslash}X}

\usepackage{listings}
\usepackage{xcolor}

\lstdefinestyle{datalog}{
    basicstyle=\footnotesize\ttfamily,
    numbers=none,
    frame=none,
    xleftmargin=8pt,
    aboveskip=6pt,
    belowskip=4pt,
    breaklines=true,
    columns=flexible,
    commentstyle=\color{gray}\itshape,
    keywordstyle=\bfseries,
    mathescape=true,
}

\lstdefinestyle{java}{
    language=Java,
    basicstyle=\footnotesize\ttfamily,
    numbers=none,
    frame=none,
    xleftmargin=8pt,
    breaklines=true,
    keywordstyle=\color{black}\bfseries,
    commentstyle=\color{gray}\itshape,
}

\usepackage{tikz}
\usepackage[most]{tcolorbox}
\usetikzlibrary{arrows.meta, positioning}

\newtcolorbox{stepbox}[2][]{
    enhanced,
    colback=white,
    colframe=gray!30,
    title=#2,
    fonttitle=\bfseries\sffamily,
    coltitle=black,
    attach boxed title to top left={xshift=3mm, yshift=-2mm},
    boxed title style={colback=gray!10, colframe=gray!30},
    drop shadow={gray!20},
    #1
}

\newcommand{\tool}{\textsc{LogicLoc}\xspace}

\newcommand{\dataset}{\textsc{KA-LogicQuery}\xspace}
\newcommand{\negset}{\textsc{KA-LogicQuery-Neg}\xspace}
\newcommand{\souffle}{Soufflé}

\begin{document}

% \title{Neurosymbolic Repo-level Code localization: Integrating Datalog Reasoning with Large Language Models}
\title{Neurosymbolic Repo-level Code Localization}

\author{Xiufeng Xu}
\authornote{Both authors contributed equally to this research.}
\email{xiufeng001@e.ntu.edu.sg}
\affiliation{%
	\institution{Nanyang Technological University}
	\country{Singapore}
}

\author{Xiuheng Wu}
\authornotemark[1]
\email{xiuheng.wu@ntu.edu.sg}
\affiliation{%
  \institution{Nanyang Technological University}
  \country{Singapore}
}

\author{Zejun Zhang}
\email{zejun.zhang@ntu.edu.sg}
\affiliation{%
	\institution{Nanyang Technological University}
	\country{Singapore}
}

\author{Yi Li}
\email{yi\_li@ntu.edu.sg}
\affiliation{%
	\institution{Nanyang Technological University}
	\country{Singapore}
}

% \renewcommand{\shortauthors}{Xu et al.}

\begin{abstract}\label{sec:abstract}
Code localization is a cornerstone of autonomous software engineering. Recent advancements have achieved impressive performance on real-world issue benchmarks. However, we identify a critical yet overlooked bias: these benchmarks are saturated with keyword references (e.g. file paths, function names), encouraging models to rely on superficial lexical matching rather than genuine structural reasoning. We term this phenomenon the \textit{Keyword Shortcut}. To address this, we formalize the challenge of Keyword-Agnostic Logical Code Localization (KA-LCL) and introduce \dataset, a diagnostic benchmark requiring structural reasoning without any naming hints. Our evaluation reveals a catastrophic performance drop of state-of-the-art approaches on \dataset, exposing their lack of deterministic reasoning capabilities. We propose \textbf{\tool}, a novel agentic framework that combines large language models with the rigorous logical reasoning of Datalog for precise localization. \tool extracts program facts from the codebase and leverages an LLM to synthesize Datalog programs, with parser-gated validation and mutation-based intermediate-rule diagnostic  feedback to ensure correctness and efficiency. The validated programs are executed by a high-performance inference engine, enabling accurate and verifiable localization in a fully automated, closed-loop workflow. Experimental results demonstrate that \tool significantly outperforms SOTA methods on \dataset while maintaining competitive performance on popular issue-driven benchmarks. Notably, \tool attains superior performance with significantly lower token consumption and faster execution by offloading structural traversal to a deterministic engine, reducing the overhead of iterative LLM inference.
\end{abstract}





\maketitle

\section{Introduction}\label{sec:intro}
The rapid evolution of artificial intelligence has fundamentally reshaped every phase of the Software Development Life Cycle (SDLC), revolutionizing developer-codebase interactions. Through natural language, developers can now gain comprehensive insight into code repositories and perform sophisticated tasks such as code refactoring, feature implementation, and defect repair. At the heart of these capabilities lies the critical primitive of \textit{code localization}. Formally, code localization is the process of precisely identifying relevant source code snippets, ranging from specific methods to complex logic blocks, within a large-scale code repository that aligns with a natural language query. As the community pivots towards autonomous software engineering agents, the efficacy of code localization has emerged as the primary bottleneck. Accurately mapping high-level functional intent to intricate implementation logic is no longer merely a retrieval challenge, but an essential prerequisite for AI systems to reliably navigate and reason about real-world software architectures.

Contemporary state-of-the-art code localization approaches, which can be broadly categorized into three paradigms including embedding-based retrieval, pipeline-guided LLM workflows, and graph-augmented agentic exploration (Details in Section~\ref{sec:related}), have reported impressive results on issue-solving benchmarks such as SWE-bench~\cite{swebench2024}. However, our investigation reveals a significant but overlooked bias that we term the \textbf{Keyword Shortcut}. Recent mainstream benchmarks are predominantly curated from GitHub issues, which usually contain clear error traces or even verbatim code snippets. Such descriptions are inherently laden with \textbf{keywords} (e.g., precise class names or unique identifiers) that act as ``cheat sheets'', allowing models to locate code snippets via surface-level lexical matching rather than genuine logical reasoning. Our preliminary diagnostic study indicates that once these identifiers are stripped away, the performance of all three approaches suffers a catastrophic decline. This stark contrast uncovers a fundamental deficiency in current systems: a profound struggle with \textbf{Keyword-Agnostic Logical Code Localization (KA-LCL)}, where models must navigate codebases without the crutch of explicit naming hints. It is crucial to distinguish KA-LCL from general issue-solving code localization: while the latter is often reducible to a \textit{semantic matching} task between query keywords and code identifiers, KA-LCL represents a higher-order \textit{structural reasoning} challenge. To illustrate this, consider a seemingly straightforward structural logical query: \textit{``Find all functions where: (1) the function has more than 15 parameters, and (2) the function is not an \_\_init\_\_ method''} (Details in Section~\ref{sec:example}). Such an keyword-agnostic logical query poses a significant hurdle for SOTA approaches. While a human developer can easily identify these patterns by traversing the program's structural logic, SOTA solutions frequently fail because they cannot rely on semantic similarity to specific identifiers. Instead, these queries necessitate a deeper understanding of code structures, where existing approaches, deprived of naming cues, prove remarkably brittle.

To systematically evaluate the limits of existing tools, we first introduce \dataset, a diagnostic benchmark specifically curated for keyword-agnostic logical code localization. Unlike widely used benchmarks that take issue statements saturated with naming hints, \dataset targets scenarios where no key entities serve as anchors. By decomposing code structures, we synthesized a series of purely logical queries that focus on code patterns. Our diagnostic evaluation reveals a precipitous performance degradation in state-of-the-art methods, uncovering a critical ``reasoning gap'' that current AI-driven localization methods have yet to bridge.

The difficulty of KA-LCL arises from two intertwined technical bottlenecks that current paradigms are ill-equipped to handle. \ding{182} The absence of lexical anchors leads to an unmanageable search space, significantly exacerbating the \textit{lost-in-the-middle} phenomenon. Modern software systems are typically massive and complex, and model-based approaches rely heavily on specific identifiers as ``pruning signals'' to filter out irrelevant modules. Without these \textit{keyword shortcuts}, models are forced to ingest a vast volume of structural context to identify potential candidates. This data deluge overwhelms the limited context window of LLMs, where critical logical patterns become submerged within long input sequences, severely impairing the system's ability to robustly access and utilize relevant structural features. \ding{183} Existing approaches lack a deterministic reasoning mechanism to navigate the intricate hierarchical dependencies of a repository-level codebase. While pipeline-guided LLM workflows and graph-augmented agentic exploration attempt to address code relationships, they often operate as probabilistic recommendation systems that generate a ranked list of likely candidates, rather than performing rigorous structural deduction. Consequently, the lack of a formal reasoning framework prevents these systems from providing either a deterministic localization or a verifiable explanation, failing to bridge the gap from heuristic-based matching to genuine repository-level logical inference.

Conceptually, code localization can be viewed as a specialized form of code search~\cite{di2023code}. However, unlike general information retrieval, programming languages possess formally defined syntax and semantics that allow source code to be precisely parsed and analyzed. This formal nature endows code with an inherent reasonability that extends beyond surface-level text. From a high-level perspective, an effective repository-level localization engine requires a robust intermediate representation (IR) to bridge the semantic gap between natural language intent and implementation logic. Such an IR must effectively encode code entities, their intricate inter-relationships, and structural hierarchies, while remaining highly interpretable and actionable for LLM-based agents.

To overcome the aforementioned limitations, we propose \tool, a novel agentic framework that synergizes the rule-based reasoning of \textbf{Datalog} with the semantic power of LLMs to achieve precise, repository-level code localization. Our framework first employs static analysis to extract a comprehensive set of \textbf{program facts} from the source code, constructing a structured IR that captures both elemental properties and relational dependencies. Upon receiving a natural language query, the LLM agent interprets the underlying functional intent and synthesizes a corresponding Datalog query. As a powerful declarative logic programming language, Datalog is uniquely suited for traversing complex structural patterns that baffle traditional retrieval methods. These queries are then executed by \textbf{Soufflé}, a high-performance reasoning engine, which performs rigorous deduction against the pre-extracted facts to infer precise code locations. We further enhance the robustness of our system with a neuro-symbolic feedback loop and mutation-based repair mechanism, enabling the LLM to iteratively refine its queries based on execution outcomes. Notably, our framework is designed to be generally applicable to any declarative logic language, including CodeQL. We chose Soufflé for several reasons: (1) it offers a clean relational syntax well-suited for LLM generation; (2) while CodeQL's query libraries are open-source, its CLI and core engine are distributed only in binary form under a separate license, which limits integration into open LLM toolchains; and (3) Soufflé supports rapid, iterative execution of partial logic rules, enabling our mutation-based diagnostic mechanism to pinpoint exactly which clause produces zero results.

Crucially, by offloading structural reasoning to a deterministic engine, \tool not only significantly reduces token consumption but also empowers the agent to provide definitive negative responses when no matches exist. This avoids the common pitfall of probabilistic systems that hallucinate potential candidates, thereby achieving a paradigm shift from heuristic-based recommendation to verifiable, high-precision localization.

\paragraph{Contributions.} Our work aims to integrate Datalog's rule-based inference engine with the advanced large language models. This framework embodies an exploration of the neuro-symbolic paradigm and hope to contribute to open science. In summary, we make the following contributions.

\begin{enumerate}
    \item We identify and formalize the \textit{Keyword Shortcut} bias in current code localization research. To address this, we introduce \dataset, a diagnostic benchmark specifically designed for Keyword-Agnostic Logical Code Localization (KA-LCL). It contains 25 high-quality purely logical queries with precise ground-truth locations, providing a rigorous testing ground for evaluating the structural reasoning capabilities of LLMs and AI agents.
    \item We proposed a novel agent-based framework for repo-level code localization that introduces program facts as an intermediate representation to capture both explicit and implicit code relationships. By synthesizing Datalog queries from natural language, \tool offloads intricate structural traversal to a high-performance deterministic reasoning engine, significantly enhancing reasoning capabilities and reducing token consumption.
    \item We implement our framework as an automated, end-to-end command-line tool. It features an iterative refinement mechanism where the LLM agent progressively generates and adjusts Datalog rules to navigate repositories. Our tool and benchmark are publicly available at: \url{https://anonymous.4open.science/r/LogicLoc-D9BE}.
    \item We conduct an extensive evaluation of \tool on both \dataset and other issue-driven benchmarks. The experimental results demonstrate that \tool significantly outperforms state-of-the-art methods in KA-LCL tasks, achieving superior precision and the capacity for verifiable localization. Furthermore, \tool maintains competitive performance on standard issue-driven benchmarks, matching SOTA levels while offering higher reliability in handling negative queries through its deterministic logic.
\end{enumerate}

\section{Background}\label{sec:bg}
This section introduces the background on program facts and Datalog that forms the symbolic reasoning component of our framework.

\subsection{Program Facts}
Program facts are a structured representation of information extracted from source code for the purpose of automated reasoning and analysis. They encode observable properties of a program, such as the existence of entities (e.g., functions, classes, variables), their attributes (e.g., names, locations, modifiers), and relations between them (e.g., containment, calls, inheritance), in a form suitable for systematic querying and inference.

Program facts are derived mechanically from source code through language-specific frontends, typically by parsing the code and traversing intermediate representations such as abstract syntax trees or control-flow graphs. Each extracted fact captures a single, well-defined aspect of the program, and together they provide a precise, machine-readable abstraction of the program’s structure and behavior. Importantly, program facts are \emph{descriptive}: they record what is present in the program, rather than how analyses should be performed.

\subsection{Datalog}

Datalog is a declarative logic programming language rooted in first-order logic and database theory. A Datalog specification consists of a set of rules that describe how new facts can be derived from existing ones.

\begin{definition}[Datalog Rule]
A Datalog rule has the form:
\begin{equation}
    R_0(t_1, \dots, t_k) \leftarrow R_1(u_1^{(1)}, \dots, u_{m_1}^{(1)}), \dots, R_n(u_1^{(n)}, \dots, u_{m_n}^{(n)})
\end{equation}
where each $R_i$ is a predicate symbol. The atom on the left-hand side is called the \emph{head}, and the atoms on the right-hand side form the \emph{body}.
Each argument $t_j$ or $u_j^{(i)}$ is either a constant or a variable.
\end{definition}

The rule is interpreted as follows: for any assignment of variables to constants that makes all body atoms simultaneously true, the corresponding instantiated head atom is also true. Variables thus serve as placeholders that allow a rule to match and relate multiple facts.

\begin{definition}[Datalog Program]
A Datalog program is a finite set of Datalog rules evaluated over a given set of ground facts. Its semantics is defined as the least fixpoint of rule application: rules are repeatedly applied to derive new ground facts until no further facts can be inferred.
\end{definition}

Datalog supports recursion and operates under a monotonic, set-based semantics, making it well suited for expressing transitive and structural properties such as reachability, dependency propagation, and hierarchical relations in program analysis.

\subsection{Program Facts in Datalog}

In program analysis, program facts are represented in Datalog as \emph{ground predicate instances}. Each predicate schema corresponds to a specific kind of program entity or relation, while each extracted program fact instantiates that schema with concrete constants derived from the source code.

For example, a predicate describing function definitions may be declared as:
    \begingroup
    \RecustomVerbatimEnvironment{MintedVerbatim}{Verbatim}{%
        commandchars=\\\{\},
        breaklines=true,
        tabsize=4,
        numbers=left, numbersep=5pt, xleftmargin=1.5em, fontsize=\footnotesize%
    }%
    \input{_minted/EDCE548FC4CFF4970B38E4A155F07A3E.highlight.minted}%
    \endgroup

An extracted function definition in the source code gives rise to a ground fact of this predicate, with all arguments bound to concrete values such as file paths, names, and line numbers.

Datalog rules operate over these ground program facts using variables to range over matching predicate instances. The body of a rule specifies patterns over existing program facts, while the head defines a new fact to be derived whenever the body is satisfied under some variable assignment. Through repeated rule application, Datalog derives higher-level program properties—such as reachability, dependency relations, or structural patterns—from the underlying set of extracted program facts.

This representation cleanly separates \emph{fact extraction}, which records concrete observations about the program, from \emph{logical inference}, which declaratively specifies how additional properties are derived.

\subsection{Motivating Example}\label{sec:example}
% To demonstrate how the aforementioned definitions are applied to code localization, we consider a practical query designed to identify functions with more than 15 parameters that are not \_\_init\_\_ methods.

\begin{figure}
    \centering
\begin{tikzpicture}[node distance=0.5em, every node/.style={inner sep=0pt}]
    \node (issue) [anchor=north] {
        \begin{stepbox}[width=\linewidth, colframe=red!50!gray, boxed title style={colback=red!10, colframe=red!50!gray}]{Question}
    \begingroup
    \RecustomVerbatimEnvironment{MintedVerbatim}{Verbatim}{%
        commandchars=\\\{\},
        breaklines=true,
        tabsize=4,
        fontsize=\footnotesize%
    }%
    \input{_minted/D66CDEA0FF4C7C85683EF78BF7259604.highlight.minted}%
    \endgroup

        \end{stepbox}
    };

    \node (code) [below=of issue] {
        \begin{stepbox}[width=\linewidth, colframe=cyan!50!gray, boxed title style={colback=cyan!10, colframe=cyan!50!gray}]{Datalog Query}
    \begingroup
    \RecustomVerbatimEnvironment{MintedVerbatim}{Verbatim}{%
        commandchars=\\\{\},
        breaklines=true,
        tabsize=4,
        numbers=left, numbersep=5pt, xleftmargin=1.5em, fontsize=\footnotesize%
    }%
    \input{_minted/EEEF6B500F7E36EEA8FE3281596A892D.highlight.minted}%
    \endgroup

        \end{stepbox}
    };

    \node (result) [below=of code] {
        \begin{stepbox}[width=\linewidth, colframe=green!50!gray, boxed title style={colback=green!10, colframe=green!50!gray}]{Query Result}
    \begingroup
    \RecustomVerbatimEnvironment{MintedVerbatim}{Verbatim}{%
        commandchars=\\\{\},
        breaklines=true,
        tabsize=4,
        fontsize=\footnotesize%
    }%
    \input{_minted/112D9585700BBDB51980E70749A13B2F.highlight.minted}%
    \endgroup

        \end{stepbox}
    };

    \draw [-{Stealth[scale=1.2]}, line width=1pt, gray!60, shorten >=-12pt, shorten <=1pt] (issue.south) -- (code.north) 
        node[pos=1.8, right=6pt, circle, fill=black, text=white, inner sep=1pt, font=\tiny, ](n1){1};
        % node[right=2pt of num1, font=\footnotesize\sffamily, text=black] {Agent};

    \draw [-{Stealth[scale=1.2]}, line width=1pt, gray!60, shorten >=-12pt, shorten <=1pt] (code.south) -- (result.north) 
        node[pos=1.8, right=6pt, circle, fill=black, text=white, inner sep=1pt, font=\tiny](n2){2};
    \node[anchor=west, at={(n1.east)}, xshift=5pt, font=\scriptsize\sffamily, text=black] {Generated by LLM};
    \node[anchor=west, at={(n2.east)}, xshift=5pt, font=\scriptsize\sffamily, text=black] {Execute by Soufflé};
\end{tikzpicture}
    \caption{A motivating exampele of a logic query}
    \label{fig:example}
    \vspace{-1em}
\end{figure}

To illustrate the query process in Datalog, we provide a motivating example. Consider a EA-LCL task that requires identifying functions in a codebase satisfying specific structural constraints: (1) the function has more than 15 parameters, and (2) the function is not an \_\_init\_\_ method.

Figure~\ref{fig:example} demonstrates a basic localization process. Given a natural language query, the Large Language Model translates the question's intention into a formal Datalog program. The generated program operates over two types of facts: \textit{Extensional Database (EDB)} facts extracted directly from source code, such as \texttt{function\_definition} containing metadata about each function's location, parameters, and context; and \textit{Intensional Database (IDB)} facts derived through logical inference, such as \texttt{LargeFunctions}. The Datalog query defines an inference rule (lines 8-11) that identifies target functions by matching against \texttt{function\_definition} facts. Irrelevant attributes are marked with underscore ``\_'' to avoid unnecessary computation, while the rule filters for functions with param\_count > 15 and excludes those named ``\_\_init\_\_''.

When executed by the Soufflé Datalog engine, the query precisely localizes two functions meeting the specified criteria: the \texttt{convolve\_fft} function with 19 parameters at line 442 in \texttt{astropy/convolution/convolve.py}, and the \texttt{\_verify\_keywords} function with 17 parameters at line 952 in \texttt{astropy/io/fits/column.py}. This example highlights the key advantages of our approach: the LLM bridges the semantic gap between natural language and formal logic, while Datalog ensures soundness and completeness of results through deductive reasoning over program facts, enabling precise localization without exhaustive manual repository traversal.
%%% Local Variables:
%%% mode: LaTeX
%%% TeX-master: "../main.tex"
%%% TeX-command-extra-options: "-no-shell-escape"
%%% End:

\section{Methodology}\label{sec:method}
In this section, we first formalize the repository-level code localization problem. We then analyze the intrinsic challenges that motivate our hybrid approach, \tool, which combines the strengths of large language models with the rigorous logical reasoning of Datalog.

Given a natural language query $q$ (e.g., a bug report or feature request) and a target codebase $\mathcal{C}$, the goal of \textbf{repository-level code localization} is to identify a list of relevant code locations $\mathcal{L}=\{l_1, l_2, \dots, l_k\}$. Effective localization bridges the gap between informal natural language and the rigid execution logic of software. We identify three primary challenges:

\begin{itemize}[leftmargin=*]
    \item \textbf{Challenge 1: Semantic and Lexical Disconnect.} There exists a non-trivial gap between the informal vocabulary of $q$ and the formal identifiers in $\mathcal{C}$. We categorize this disconnect into three progressive layers: (1) \textit{Keyword Absense}, where a query lacks any direct textual anchors present in the code; (2) \textit{Latent Semantic Mapping}, where high-level task descriptions (e.g., \texttt{login failure}) lack direct textual overlap with low-level implementation entities (e.g., \texttt{AuthManager} or \texttt{ValidateToken}); (3) \textit{Lexical Divergence}, where developers use synonyms or abbreviations (e.g., \texttt{fqn} for \texttt{fullyQualifiedName}) that elude exact match search.

    \item \textbf{Challenge 2: Non-local Structural Dependencies.} Relevant code is often not directly mentioned in queries but connected through dependencies or calls. In modern software, even a single feature may be implemented across multiple files. For example, identifying \textit{password validation} logic may require traversing complex call chains and data flows scattered across authentication modules and database layers. Existing pipeline-based tools rely on local directory traversal and fail to capture these deep, cross-file relational dependencies.
    
    \item \textbf{Challenge 3: Complex Logical Pattern Constraints.} Questions may contain logical pattern requirements that candidate code must satisfy. Certain localization tasks are defined by structural patterns rather than keywords. For example, \textit{``identifying the conditional statement that raises three distinct error types in different branches''} requires satisfying specific logical constraints defined by the language's syntax and semantics. Such patterns are difficult to locate by simply retrieving class or function information; it requires a more comprehensive understanding of the codebase and reasoning ability.
\end{itemize}

To address these challenges, we propose \tool, a synergy between LLMs and Datalog. Our intuition is that LLMs excel at resolving Challenge 1 by translating vague natural language intents into structured requirements, while Datalog provides the relational and reasoning power to solve Challenge 2 and 3 by performing exhaustive, sound traversal over codebase. 
In our framework, we represent the codebase $\mathcal{C}$ as a set of pre-extracted program facts $\mathcal{F}$, where $\mathcal{F}$ captures both the source entities and the structural relations (e.g., call graphs, inheritance). Each location $l_i \in \mathcal{L}$ corresponds to a specific level of granularity, such as a file, module, or function, that is essential for resolving the query $q$. Figure~\ref{fig:arch} illustrates the overview of the framework of \tool. Our framework operates in two stages. First, an offline program fact extraction stage analyzes the codebase to build a structured knowledge base. Second, an automated agent execution stage leverages these extracted facts to perform code localization by synthesizing high-quality Datalog queries.

% \subsection{Neurosymbolic Architecture Overview}
% Our solution integrates Large Language Models with Datalog-based program querying to bridge the semantic gap while ensuring structural completeness. Figure~\ref{fig:arch} illustrates the overview of the framework of \tool. Our framework operates in two stages. First, an offline program fact extraction stage analyzes the codebase to build a structured knowledge base. Second, an automated AI agent execution stage leverages these extracted facts to perform code localization.

\subsection{Program Facts Extraction}
In this step, we will discuss the details of extracting program facts from a given repository.
We extract program facts through a modular pipeline that separates source discovery, structural parsing, and fact emission. First, we enumerate source files under the repository root using configurable include/exclude patterns that respect version-control ignores, build artifacts, and generated code. Language frontend then analyzes its files using the most suitable intermediate representation. For Python, we parse source code into an abstract syntax tree and emit facts during traversal. The frontend produces Datalog facts annotated with precise source locations and stable identifiers, enabling traceability and incremental updates.

% For Java, we utilize Soot~\cite{github:soot} to generate an internal program representation and enable static analysis, serving as the basis for fact extraction.

From these frontends, we emit facts describing program entities (files, modules, functions, classes, variables) and relations (containment, inheritance, import/use, call, and reference edges), together with optional control flow and data flow information when available. This representation directly addresses Challenge 2 (non-local structural dependencies) by making cross-file interactions explicit and queryable~\cite{wu2021diffbase}. Instead of relying on local directory traversal, localization can now operate over global dependency graphs, enabling the identification of code relevant to a feature even when it is scattered across multiple modules and layers.

Meanwhile, our facts enables expressive logical pattern matching, which helps solving the Challenge 3 (complex logical pattern constraints). Structural requirements, such as the presence of specific exception patterns, or multi-step call sequences, can be formulated as Datalog queries over extracted facts. This allows localization tasks defined by program structure rather than surface keywords to be handled systematically, without requiring the LLM to reason over raw source code.

%We also customize fact schemas across languages to reflect language-specific constructs and frontend capabilities. It provides a normalized structural interface that allows LLMs to operate over rich, explicit program representations rather than plain text alone. In our hybrid setting, semantic bridging between natural-language queries and program intent is primarily handled by LLMs, while our fact extraction ensures that once candidate concepts are identified, they can be grounded in precise, analyzable code structures.

\subsection{Agentic Workflow}
This section elaborates on our agent-based workflow for automated repository-level code localization, which builds upon the program facts constructed offline.

Our end-to-end agent is designed to accept natural language queries and return precise code locations. Unlike approaches that provide a fixed top-$n$ list of candidates, our system outputs a dynamic set of potential locations to maximize precision. To mitigate hallucinations and enhance abstention ability, we decouple reasoning from generation. A deterministic engine handles inference while the LLM functions as a coordinator for query analysis and result calibration. The agent is equipped with three basic tools: ``\texttt{exec\_dl}'' for executing Datalog programs, and ``\texttt{get\_file\_contents}'' along with ``\texttt{get\_sources}'' for retrieving source code and specific line ranges.

\subsubsection{Query Analysis and Information Resolution}
The workflow begins with performing a preliminary analysis to extract the core technical concepts and structural elements. By identifying program entities like specific file paths and module names, and core structure descriptions, the agent establishes an internal context. This preparatory step provides the necessary predicates and constraints for the subsequent synthesis of Datalog programs.

% \subsubsection{Syntax Correction and Intermediate-rule Diagnosis}
\subsubsection{Synthesize-Check-Refine Loop}
\begin{figure}[t]
	\centering
	\includegraphics[width=\textwidth]{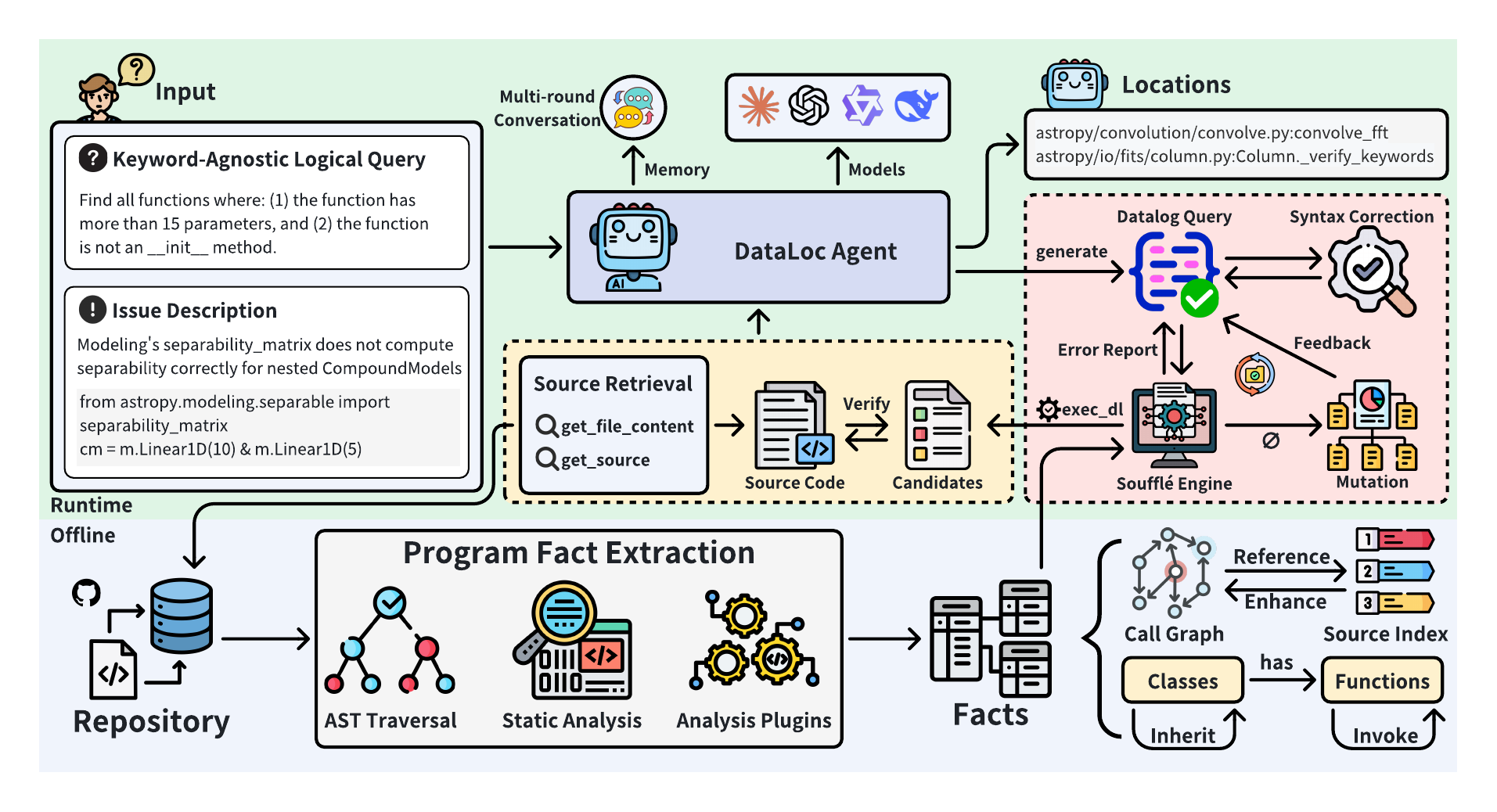}
	\caption{Overview of \tool framework}
	\label{fig:arch}
\end{figure}

As shown in the red box in the Figure~\ref{fig:arch}, before execution, \tool follows a \textit{synthesize-check-refine} loop to mitigate the impact of hallucinations and improve the executability of LLM-generated Datalog programs. It mainly contains two critical, feedback-driven phases (Details in Section~\ref{sec:method:val} and \ref{sec:method:int}): (1) \textit{Syntax correction}. Each synthesized program will undergo a parser-gated validation to ensure syntactic well-formedness. Our workflow adopts a \emph{best-effort repair, then fallback} strategy: unambiguous cases are handled via conservative rule-based fixes, revalidated with \souffle’s parser, and all other cases return error feedback to the LLM. (2) \textit{Intermediate-rule diagnosis}. We instrument the program to track row counts of intermediate relations, thereby identifying rules that produce empty results. Using controlled, mutation-based probing, we differentiate fragile-empty relations caused by over-constraints from stable-empty relations that reflect inherent dataset properties. These feedbacks help the model refine the generated program and ensure its quality before it reaches the inference engine. Validated programs are then executed through \texttt{exec\_dl} tool to generate a list of candidates. Excessive locations are treated as failures, triggering refinement with stricter constraints to reduce noise and improve precision.

% Programs undergo an initial syntactic parsing where simple errors are auto-corrected and ambiguous ones trigger an immediate return to the LLM. Survivors then proceed to semantic checks targeting Soufflé-specific constraints, with feedback provided for uncertain cases.

% our workflow integrates two critical, feedback-driven phases: parser-gated validation and mutation-based intermediate-rules diagnosis. 

% this process follows a ``synthesize-check-refine'' loop, as depicted in :

% \begin{enumerate}
%     \item \textit{Parser-Gated Validation.} Before execution, synthesized queries undergo a parser-gated validation workflow to ensure syntactic well-formedness. If failures occur, the system applies a \textit{best-effort repair} strategy, using rule-based fixes for unambiguous issues (e.g., reserved keyword renaming) or providing raw parser diagnostics back to the LLM for global restructuring.
%     \item \textit{Semantic Correctness Checking.} After passing the parser check, the workflow applies a set of semantic rules to rectify common logical misconceptions, such as inverted argument semantics in built-in predicates (e.g., contains). This ensures the query is not only syntactically correct but also logically sound before it reaches the execution engine.
%     \item \textit{Execution and Result Refinement.} Validated queries are executed via the \texttt{exec\_dl} tool. If the query returns an excessive number of results, the agent treats it as a failure and refines the logic with stricter constraints to minimize noise and improve precision.
% \end{enumerate}

\subsubsection{Context Retrieval and Final Verification}
As depicted in the yellow box in Figure~\ref{fig:arch}, once a set of potential code locations has been determined, the agent retrieves the relevant code snippets using \texttt{get\_sources} or \texttt{get\_file\_contents} and conducts a verification against the original query. These verified locations are then returned in a standardized format (e.g., \texttt{FILE\_PATH:CLASS.METHOD}). Although the agent may undergo multiple internal reasoning iterations, the user experience is streamlined into a single step: submitting a query and receiving a list of locations. This fully automated closed-loop design ensures both usability and seamless integration into production development environments.

% \begin{table}[htbp]
%     \centering
%     \caption{Available Tools for Agent}
%     \label{tab:available-tools}
%     % \vspace{2mm} 
%     \small
%     \begin{tabularx}{0.9\textwidth}{@{} l X @{}}
%         \toprule
%         \textbf{Tool Name} & \textbf{Description} \\
%         \midrule
%         \texttt{exec\_dl} & Execute Datalog programs (schema depends on detected language). \\
%         % \addlinespace
%         \texttt{get\_file\_contents} & Get complete source file contents using appropriate identifiers. \\
%         % \addlinespace
%         \texttt{get\_sources} & Get multiple specific line ranges from source files. \\
%         \bottomrule
%     \end{tabularx}
% \end{table}

\subsection{Program Repair for LLM Generated Datalog}
\label{sec:method:val}
In practice, LLM-generated Datalog programs (the dialect used by \souffle, in our case) often contain syntactic and semantic errors, especially when the model is not fine-tuned for Datalog programming.

Before executing an LLM-generated Datalog program, we enforce a \emph{parser-gated validation} workflow to ensure that only syntactically well-formed programs reach later stages of the pipeline. This design is motivated by the observation that some failures are caused by superficial syntactic issues that can be repaired deterministically, while more complex parse failures often require global restructuring that is better handled by the LLM. Our workflow therefore follows a \emph{best-effort repair, then fallback} strategy: we apply conservative rule-based fixes when the repair is unambiguous, re-check the program using \souffle's parser, and otherwise return error feedback to the LLM.

We invoke a lightweight parser helper (based on \souffle's parsing frontend) to validate the generated program. If parsing fails, we first attempt a small set of mechanical rewrite rules targeting high-frequency issues. For example, LLMs frequently introduce naming collisions by using reserved or special identifiers as variable names. E.g., using \texttt{count} as a variable name while it is a reserved keyword in \souffle. Such cases can be fixed locally via deterministic renaming.
%We similarly normalize other local parse hazards (e.g., missing terminators, malformed aggregate punctuation) when the repair is syntactically and semantically safe.

However, not all parser errors admit a reliable deterministic repair.
%Complex failures,s uch as malformed rule structure, severely mismatched parentheses, or directive misuse intertwined with rule bodies, often have multiple plausible fixes and may require the program to be re-synthesized rather than patched.
In these cases, we do not attempt speculative transformations. Instead, we return the raw parser diagnostics (optionally augmented with concise hints) to the LLM, allowing the model to revise the program directly.

Any program that does not pass the parser check is rejected and never proceeds to semantic validation or execution. Only after the program passes syntactic validation (either initially or after rule-based fixes) do we apply semantic rule checking and subsequent execution. Then we apply a set of semantic correctness checks that encode \souffle-specific usage rules and common domain conventions observed in LLM-generated programs. These checks target misconceptions that LLMs frequently exhibit when generating Datalog program.
%This staged design prevents cascading failures in later steps and ensures that semantic checking operates over a well-defined abstract syntax.

%Relying solely on iterative prompting or post-hoc error messages from the Datalog engine can lead to repeated failures and excessive tool calls. To address this issue, we introduce a repair component that automatically validates and repairs LLM-generated Datalog programs before sending it to the \souffle{} engine.

%We first employ \souffle's parser to perform strict syntactic validation on the generated program. Many syntactic issues can be easil
%Unlike treating the engine as a black box that only reports errors after a failed execution, we explicitly surface parse-time diagnostics and use them to drive targeted rewrites. Common syntactic issues include missing rule terminators, malformed aggregates, unmatched parentheses, and incorrect use of directives such as .decl, .input, and .output.
%For high-confidence cases, we apply deterministic fixes (e.g., normalizing aggregate syntax, inserting missing delimiters, or correcting directive placement). Programs that cannot be repaired unambiguously at this stage are passed through unchanged but annotated with structured diagnostics for downstream hint generation.

A representative example is the use of string containment constraints. \souffle{} provides a constraint function
\texttt{contains(sub:symbol,full:symbol)}, which is defined such that the second argument must contain the first (i.e., full includes sub). However, we observe that LLMs frequently invert this positional relationship by producing atoms like \texttt{contains(content,"keyword")} instead of the correct \texttt{contains("keyword",content)}. Since such inversion conforms to both syntax and type specifications, it leads to silent failure or empty results that are difficult for LLM to self-correct, even with multiple iterations of try and feedback.
% This error does not trigger any syntax or type error, produces a well-formed but logically inverted condition, and leads to silent failure or empty results that are difficult for LLM to realize, even with multiple iterations of try and feedback.

Prompt techniques such as few-shot learning cannot effectively eliminate this kind of error. Therefore, we construct a library of semantic rules that check the correct usage of built-in predicates with non-commutative argument semantics and other similar constraints. These repairs are applied only when the transformation is high-confidence and semantics-preserving. When uncertain, the checker records the issue for subsequent feedback to the LLM rather than applying a blind fix, thereby guiding the refinement in subsequent iterations.
We evaluate the contribution of this mechanism through an ablation study in~\cref{sec:ablation}, demonstrating its effectiveness in improving synthesis quality. 

% If uncertainty exists, the checker refrains from modifying the program and instead records the issue for later feedback to the LLM, guiding it to generate better results in subsequent iterations.

\subsection{Diagnosing Intermediate Rules via Conservative Mutation Analysis}
\label{sec:method:int}
\begin{figure}[t]
	\centering
	\includegraphics[width=\textwidth]{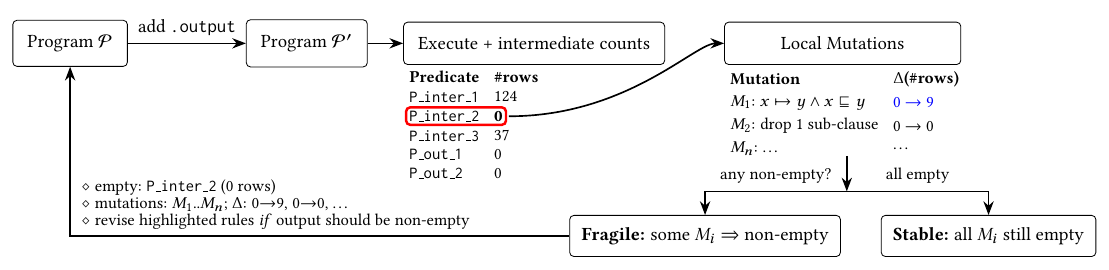}
	\caption{Intermediate rules mutation and feedback}
	\label{fig:mutate}
\end{figure}
We introduce an intermediate-rule diagnostic and mutation-based feedback mechanism to improve both the efficiency and effectiveness of LLM-based Datalog synthesis. As illustrated in~\cref{fig:mutate}, instead of evaluating a generated program solely by its final output, we instrument execution to collect row counts for intermediate relations and identify rules whose derived relations are empty. In practice, empty intermediate relations frequently indicate overly restrictive constraints, mismatched join keys, or incorrect predicate usage, and therefore serve as a useful signal for localizing potential errors in synthesized programs.

An empty relation is not inherently incorrect: depending on the user’s constraints and the underlying dataset, the semantically correct result may legitimately be the empty set. To avoid forcing spurious revisions or encouraging the model to hallucinate evidence, our approach explicitly distinguishes between fragile and stable emptiness through controlled diagnostic probing.

\subsubsection{Mutation Operators}
\label{sec:mutation-operators}

We define a fixed catalog of five diagnostic mutation operators, organized
into two families. Each operator is \emph{structure-preserving}: it
modifies only the body of a rule while leaving the head (and therefore the
relation schema) unchanged. Mutations are purely diagnostic---they are
never used as candidate replacements for the final query.

\paragraph{Family~1: String-relaxation mutations.}
These operators relax exact string equality to approximate matching,
targeting the most common error class in LLM-generated Datalog programs
over code-analysis fact bases: over-specified string literals. Because the
LLM must guess identifier formats (fully qualified names, file paths,
method signatures) without seeing the data, exact-match constraints
frequently fail even when the logical intent is correct.

\begin{enumerate}[label=\textbf{M\arabic*}.,leftmargin=*]

\item \textbf{\textsc{Contains-Literal} (substring relaxation).}
For every string literal \texttt{"s"} appearing in the rule body, the
mutation replaces it with a fresh variable~$V_i$ and appends a
\texttt{contains("s", $V_i$)} atom to the body, relaxing exact equality to
substring containment. For example, the atom
\texttt{IsMethod(\_, "getValue", \ldots)} becomes
\texttt{IsMethod(\_, MUTVAR1, \ldots)} and \texttt{contains("getValue", MUTVAR1)}.

\item \textbf{\textsc{Match-Literal} (prefix relaxation).}
Analogous to \textsc{Contains-Literal}, but uses
\texttt{match}\texttt{("$s$.*", $V_i$)} instead of
\texttt{contains}. This tests whether the original literal matches as a
\emph{prefix} of the actual data, diagnosing a different failure mode:
the literal may be correct as a prefix but actual values carry trailing
context (e.g., type signatures).
\end{enumerate}

\paragraph{Family~2: Single-constraint ablation mutations.}
Rather than modifying a constraint's matching semantics, these operators
\emph{remove} exactly one body atom from a rule, producing a separate
diagnostic variant for each eligible atom. This one-at-a-time strategy
provides fine-grained localization: by observing which single removal
causes rows to appear, the system pinpoints the specific constraint
responsible for emptiness.

\begin{enumerate}[label=\textbf{M\arabic*}.,leftmargin=*,start=3]

\item \textbf{\textsc{Drop-Single-Atom} (general single-constraint ablation).}
For each body atom~$a_i$ in rules defining the target relation, the
system generates a variant with~$a_i$ removed and all other atoms
retained. If a rule has $k$ body atoms ($k \geq 2$), this produces up to
$k$ variants per rule. For example, given the rule:
\begin{lstlisting}[language=Prolog,basicstyle=\small\ttfamily]
R(x, y) :- A(x, y), B(y, z), contains("foo", x).
\end{lstlisting}
three variants are generated, each dropping one of
\texttt{A(x,y)}, \texttt{B(y,z)}, or \texttt{contains("foo",x)}.
If dropping \texttt{B(y,z)} is the only variant that yields rows, the join
on \texttt{B} is identified as the likely error site. To control
computational cost, the total number of single-drop variants per relation
is capped at a configurable limit (default~3).

\item \textbf{\textsc{Drop-Single-Contains} (targeted string-filter ablation).}
A specialization of \textsc{Drop-Single-Atom} restricted to
\texttt{contains(\ldots)} atoms. Removing one \texttt{contains} filter at a time isolates
whether a specific filter (rather than the combination of all filters) is
the cause of emptiness.

\item \textbf{\textsc{Drop-Single-Negation} (targeted negation ablation).}
A specialization restricted to negated atoms (\texttt{!pred(\ldots)}).
LLMs sometimes introduce incorrect negation constraints---e.g., excluding
a class from results when the intent was to include it. Removing one
negated atom at a time tests whether any individual negation is overly
restrictive.
\end{enumerate}

\subsubsection{Mutation Selection and Applicability}
\label{sec:mutation-selection}

Not every mutation operator is applicable to every rule. The system
determines applicability via lightweight syntactic checks before execution:

\begin{itemize}
\item \textsc{Contains-Literal} and \textsc{Match-Literal} require
at least one string literal (a quoted string) in the rule body.

\item \textsc{Drop-Single-Atom} requires at least two body atoms (removing
the sole atom would produce an unconditional fact with potentially
ungrounded variables).

\item \textsc{Drop-Single-Contains} requires at least one
\texttt{contains(\ldots)} atom in a rule with $\geq 2$ body atoms.

\item \textsc{Drop-Single-Negation} requires at least one negated atom
(\texttt{!pred(\ldots)}) in a rule with $\geq 2$ body atoms.

%\item \textsc{Drop-All-Contains} requires at least one
%\texttt{contains(\ldots)} atom.
\end{itemize}

\subsubsection{Classifying Empty Relations and Constructing Feedback}
\label{sec:mutation-classification}

After all applicable mutations have been executed for a zero-row
relation~$R$, the system classifies the emptiness and constructs targeted
feedback:

\paragraph{Fragile-empty.}
If at least one successfully executed mutation produces a non-empty result
for~$R$ (i.e., $\text{rows}(R_{\text{mutated}}) > 0$), the relation is
classified as \emph{fragile-empty}. This signals that the original rule is
likely over-constrained or mis-specified: a small, conservative relaxation
sufficed to surface rows, suggesting the original constraints were close
to---but not quite matching---the actual data. The feedback returned to the
LLM includes:

\begin{itemize}
\item The name of the affected relation and its zero row count in the
original execution.
\item For each mutation that produced rows: the mutation operator identity,
the specific constraint that was relaxed or removed (for single-constraint
ablation mutations, this identifies the exact atom), the rewritten rule(s),
the resulting row count, and (optionally) a small sample of newly surfaced
tuples (capped at a configurable limit, default~5). The sample helps the
LLM see the actual data format and adjust accordingly.
\item For each mutation that remained empty or failed: the operator
identity and its outcome, providing a complete diagnostic picture.
\end{itemize}

\paragraph{Stable-empty.}
If \emph{all} applicable mutations either remain empty or fail, the
relation is classified as \emph{stable-empty}. In this case, emptiness may
reflect a genuine property of the dataset under the stated constraints
rather than a synthesis error. To remain conservative, we do not pressure
the model to introduce relaxations; instead, we report that the empty
result appears robust under diagnostic probing and encourage the model to
either preserve the current semantics or emit auxiliary diagnostic outputs
(e.g., intermediate row counts for upstream relations) to further
investigate.

\paragraph{Mutations as probes, not patches.}
A critical design principle is that results produced by mutated programs
are \textbf{never} used as final answers. Mutations serve exclusively as
execution-guided diagnostic probes. The LLM receives the mutation feedback
as part of the tool response alongside the original (empty) query results,
and decides autonomously whether and how to revise its program. This
separation ensures that the system does not silently weaken the user's
intended query semantics: it provides evidence to support the model's
reasoning while leaving the revision decision to the model.

\section{Evaluation}\label{sec:eval}

To evaluate the effectiveness and practicality of our approach at repository-level, we design the following research questions:
\begin{enumerate}
	\item \textbf{RQ1:} How effective is \tool in keyword-agnostic logical code localization?
	\item \textbf{RQ2:} How effective is \tool for issue-based code localization?
	\item \textbf{RQ3:} How efficient is \tool compared to baselines?
    \item \textbf{RQ4:} How does each component of \tool contribute to its performance?
\end{enumerate}

\subsection{Experiment Setup}

\subsubsection{Benchmarks.} 
We evaluate code localization performance on three Python-based benchmarks, covering both complex logical reasoning challenges and industrial issue-resolution tasks.

\textbf{SWE-bench Lite}~\cite{swebench2024}. A carefully curated and widely recognized subset from the full SWE-bench for more efficient and cost-effective evaluation of autonomous issue-solving capabilities. It consists of real-world GitHub issues with repository metadata and ground-truth patch locations. Following Suresh et al.~\cite{suresh2024cornstack}, we retained 274 of 300 original instances where patches modify existing functions or classes. We intentionally excluded instances introducing code corresponding to new functions or import statements to focus the evaluation on code localization within existing structures.

\textbf{\dataset} (Ours). To evaluate the capability of localization approaches in keyword-agnostic logical code localization, we constructed \dataset, a diagnostic benchmark comprising 225 high-quality logic-intensive queries across 9 projects. As illustrated in Table~\ref{tab:code_dimensions}, each query is formulated as a composite logical proposition by integrating code features across multiple dimensions. By combining structural granularity (e.g., classes, methods) with behavioral attributes (e.g., control flow, exception handling) and code metrics (e.g., inheritance depth and branch count), the feature space admits a large number of possible compositions. We selected 25 representative combinations to serve as our query set, each applied across 9 repositories at the base commit versions sampled from SWE-bench Lite, with minimal adjustments to the query itself to ensure that at least one valid ground-truth location exists in the corresponding codebase. Structural reasoning is a prerequisite for bridging the semantic gap: if an agent cannot even resolve explicit structural constraints, it certainly cannot handle latent semantic mappings where such patterns are implicit. \dataset tests a necessary condition for genuine code understanding.

To establish ground truth, we ran multiple state-of-the-art models (e.g., Claude-4.5-Opus, GPT-5.2) in the agent mode of advanced AIDEs (\texttt{Cursor} and \texttt{GitHub Copilot}) to generate candidate locations, then aggregated all returned results to maximize coverage. Two authors independently validated each candidate against the query constraints, resolving disagreements through discussion grounded in the formal Python syntax specification.

\dataset serves as a critical complement to issue-based benchmarks for localization task. In practice, issues are one of the most important channels for error feedback between users and maintainers. To facilitate debugging, those issue descriptions often provide sufficient information and clear keywords as cues to help maintainers better locate faults, such as accurate file paths, function identifiers, or even specific code snippets. Our analysis of SWE-bench Lite instances reveals that for over 50\% of cases, the issue description explicitly mentions identifying information of the ground-truth location (e.g., file names, class names, or function names). Such \textit{keyword shortcut} enables models to succeed via simple lexical matching (e.g. grep) or embedding-based retrieval, without requiring genuine understanding and reasoning over the codebase. This undermines the validity of localization performance evaluations. Moreover, LLM-assisted development shifts the codebase interaction toward intent-based question answering, allowing developers to query repositories using natural language. However, for developers unfamiliar with a given repository, they typically cannot use precise identifiers and instead tend to express their search intent through high-level behavioral pattern descriptions or abstract logical structures.

\textbf{\negset} (Ours). \negset is a variant of \dataset designed to evaluate a system's abstention capability when no valid location meets the query. Verifying a model's ability to return an empty set does not require evaluation at this full scale of the original dataset. Therefore, we construct \negset by sampling a single version from one repository and intentionally modifying its queries to ensure their ground-truth sets are empty. Current methods often adopt top-$n$ ranking to maximize recall, but ideal robust localization requires the ability to provide ascertained answers and explicitly avoid false positives. Such a ``refusal'' mechanism is a critical metric for ensuring the reliability of autonomous agents in production environments.

\begin{table}[htbp]
\centering
\caption{Taxonomy of Python Code Dimensions and Representative Elements}
\label{tab:code_dimensions}
\small
\renewcommand{\arraystretch}{1.2}
\begin{tabularx}{\textwidth}{lX}
\toprule
\textbf{Query Dimensions} & \textbf{Examples / Typical Elements} \\ \midrule
Code Structure & Functions, Methods, Classes, Modules, Decorators \\
Control Flow & Conditional (\texttt{if-elif-else}), Iteration (\texttt{for}, \texttt{while}), Context Management (\texttt{with}) \\
Condition Logic & Comparison (\texttt{==}, \texttt{>}), Identity (\texttt{is}), Membership (\texttt{in}), Type Checks (\texttt{isinstance}), Logical Operators (\texttt{and}, \texttt{or}, \texttt{not}), Early Exit (\texttt{return}, \texttt{break}, \texttt{continue}) \\
Data Structure & Built-in Collections (\texttt{list}, \texttt{dict}, \texttt{set}), Primitive Types (\texttt{int}, \texttt{str}) \\
Function Signatures & Default Values, Variadic Parameters (\texttt{*args}, \texttt{**kwargs}), Type Annotations \\
Exception Handling & Exception Propagation (\texttt{try-except-finally}), Exceptions (\texttt{TypeError}) \\
Code Metrics & Nesting Depth, Inheritance Depth, Assertion Count, Branch count \\ \bottomrule
\end{tabularx}
\end{table}

\subsubsection{Baselines.} To assess \tool, we select four state-of-the-art baselines representing three distinct technical paradigms: embedding-based, pipeline-based, and agent-based approaches:
\begin{enumerate}[leftmargin=*]
    \item \textbf{SweRank}~\cite{reddy2025swerank} (\textit{Embedding-based}): It utilizes a retrieve-and-rerank architecture to identify issue locations. It employs SWERankEmbed (137M/7B parameters) to perform initial retrieval and SWERankLLM (7B/32B parameters) to rerank the results.
    \item \textbf{Agentless}~\cite{xia2024agentless} (\textit{Pipeline-based}): This approaches employs a hierarchical filtering strategy within a procedural workflow. It progressively prunes the search space from the file level down to specific classes or functions, utilizing an LLM to rank and select candidates at each stage.
    \item \textbf{LocAgent}~\cite{chen-etal-2025-locagent} (\textit{Agent-based}): It constructs a graph-based representation and sparse indexes of the project and enable an autonomous agent to perform iterative, tool-assisted retrieval.
    \item \textbf{CoSIL}~\cite{jiang2025cosil} (\textit{Agent-based}): This framework focuses on structural dependency traversal through call graphs to identity implicit locations via iterative exploration. It incorporates pruning to maintain context efficiency and restrict the search to high-relevance execution paths.
    % \item \textbf{Orca Loca~\cite{yu2025orcalocallmagentframework}:} Integrates priority-based scheduling, action decomposition with relevance scoring, and distance-aware context pruning. By optimizing the synergy between agentic reasoning and precise retrieval, it effectively navigates complex repositories to resolve the suboptimality of current search mechanisms.
\end{enumerate}

\subsection{Metrics}
We evaluate localization performance at three granularity: \textit{file, module, and function}. Let $Q$ denote the set of query instances, $G_q$ the set of ground-truth locations for query $q \in Q$, and $\mathcal{P}_q$ the set of predicted locations inferred by our agent workflow. $\mathbf{1}(\cdot)$ denotes the \textbf{indicator function}, which equals 1 if the logical condition holds and 0 otherwise. We adopt the following six metrics:

\begin{enumerate}[leftmargin=*, label=\textbf{M\arabic*.}]
    \item \textbf{Accuracy@k (ACC@k):} It measures the ability to achieve full coverage, where a success requires all ground-truth locations to be present within the top-$k$ predicted locations. When $k$ equals the length of the prediction set, this metric becomes the \textbf{Success Rate (SR)}:
    \begin{equation}
        Acc@k = \frac{1}{|Q|} \sum_{q \in Q} \mathbf{1}(G_q \subseteq \mathcal{P}_{q,k})
    \end{equation}

    \item \textbf{Recall (REC):} It represents the proportion of ground-truth locations successfully captured by the predicted set $\mathcal{P}_q$:
    \begin{equation}
        Rec@k = \frac{1}{|Q|} \sum_{q \in Q} \frac{|G_q \cap \mathcal{P}_{q,k}|}{|G_q|}
    \end{equation}

    \item \textbf{Precision (PRE):} This metric penalizes overprediction by calculating the fraction of predicted locations that are correct:
    \begin{equation}
        Pre = \frac{1}{|Q|} \sum_{q \in Q} \frac{|G_q \cap \mathcal{P}_q|}{|\mathcal{P}_q|}
    \end{equation}

    \item \textbf{Average Jaccard Similarity (AJS):} It quantifies the overlap between the predicted and ground-truth sets, which penalizes both missing targets and redundant predictions:
    \begin{equation}
        AJS = \frac{1}{|Q|} \sum_{q \in Q} \frac{|G_q \cap \mathcal{P}_q|}{|G_q \cup \mathcal{P}_q|}
    \end{equation}

    \item \textbf{Perfect Location Rate (PLR):} The most Stringent metric, measuring the ratio of instances where the predicted set $\mathcal{P}_q$ exactly matches the ground-truth set $G_q$. A PLR of 1.0 indicates perfect localization without any extraneous noise (i.e., $AJS = 1.0$):
    \begin{equation}
        PLR = \frac{1}{|Q|} \sum_{q \in Q} \mathbf{1}(\mathcal{P}_q = G_q)
    \end{equation}

    \item \textbf{Hit Rate (HR):} The most lenient metric, measuring the ratio of instances where the predicted set $\mathcal{P}_q$ provides at least one correct location:
    \begin{equation}
        HR = \frac{1}{|Q|} \sum_{q \in Q} \mathbf{1}(\mathcal{P}_{q} \cap G_q \neq \emptyset)
    \end{equation}
\end{enumerate}

\subsubsection{Implementation and environment.} All experiments were conducted on a server equipped with an Intel Xeon Silver 4216 CPU (2.10 GHz) and 62 GB RAM, running Ubuntu 22.04.5 LTS. Our framework was implemented using Python 3.12.11 and the Soufflé 2.4 Datalog engine. To evaluate \tool, we accessed \texttt{gpt-4o-20240513} via OpenAI’s API, \texttt{claude-3-5-sonnet-20241022} through AWS Bedrock services, and \texttt{Qwen3-Max} via Alibaba Cloud Service. For all models, the temperature was set to 0.3, and the number of max iterations was fixed at 20. For baseline comparisons, we instantiated runtime environments according to their respective official specifications and dependency requirements to ensure a fair evaluation.

\subsection{Results}

\subsubsection{Effectiveness for logic query}

% To evaluate the effectiveness of \tool, we conducted extensive experiments on the \dataset dataset across three granularities: File, Module, and Function levels. We compared \tool against four state-of-the-art (SOTA) baselines: SweRank, Agentless, LocAgent, and CoSIL. The comprehensive results are presented in Table~\ref{tab:comparison_lq}.

As summarized in Table~\ref{tab:comparison_lq}, \tool achieves a decisive lead over all baselines across all metrics and granularities. At the file level, \tool reaches a Precision (PRE) of 73.35\% and a Success Rate (SR) of 68.44\%, which is significantly higher than the baseline methods. Additionally, while all baselines achieve near-zero Perfect Location Rate(PLR), \tool attains a PLR of 48.44\%. Even at the finest function-level granularity, \tool still maintains a PLR of 38.27\%, while all baselines drop to 0\%. This highlights the unique advantage of our framework in capturing code structure and reasoning capabilities in assisting precise localization. Furthermore, \tool consistently achieves exceptional performance across different foundation models, demonstrating strong adaptability. By switching the underlying model from Claude-3.5 to Qwen3-Max, the framework yields further performance gains, suggesting that our framework can scale its localization effectiveness with the advancing capabilities of underlying LLMs. 

Even with the most lenient metric, Hit Rate (HR), which only requires at least one correct location, baseline performance drops quickly as the granularity shifts from file-level to module-level and function-level. Other metrics even approach zero. It indicates that, when deprived of explicit keywords to narrow down the search space and forced into deep tracing, they struggle to maintain meaningful localization capability. In contrast, \tool exhibits remarkable resilience, achieving a high HR of around 90\% across all granularities. This robustness proves that \tool's success is not a byproduct of a coarse search space but is driven by rigorous, logic-based reasoning.

Baselines typically rely on top-$n$ recommendations to increase the probability of covering relevant locations, but this strategy is inherently a compromise rather than an optimal solution. An effective code localization tool should return results that precisely satisfy the query constraints, since the true number of relevant locations varies across tasks and is not predetermined. To evaluate this capability, we introduce two additional metrics: Average Jaccard Similarity (AJS) and Perfect Localization Rate (PLR). AJS penalizes both false positives and false negatives, while PLR represents the most stringent criterion, requiring the predicted set to exactly match the ground truth (i.e., achieving 100\% AJS). For instance, while LocAgent (Claude-3.5) achieves a 54.67\% hit rate at the file level, its AJS is only 8.73\%, indicating that true positives are diluted within an inflated candidate set containing substantial noise. By comparison, our approach consistently maintains high AJS scores, reflecting greater precision in returning constraint-satisfying results without extraneous recommendations. This precision is important for industrial deployment, as it reduces the validation overhead for developers or downstream automated agents, improving the efficiency of maintenance workflows.

Our investigation of SWE-bench Lite shows that most issues are highly localized, involving an average of only 1.15 code changes (computed as the total number of ground-truth locations divided by the total number of instances). This sparsity raises a key question: do existing tools truly pinpoint root causes, or do they merely rely on high-probability guessing within a narrow search space? To examine this, we use \negset to evaluate whether tools can recognize when no valid location exists. By modifying constraints, we deliberately created a mismatch between the issue description and the codebase, such that the original ground-truth locations are no longer valid. In this setting, the only correct output is a clear ``no match found''. Unfortunately, all SOTA baselines suffer from a compulsion to guess. They persistently return top-$n$ recommendations even when query prerequisites are not met. This over-eager behavior proves harmful in practice, as confident yet wrong targets mislead downstream agents, wasting computational resources, and risk introducing regression bugs. These findings suggest that the strong performance reported by existing baselines is partially inflated by their recommendation-centric design, which lacks true localization rationale. Notably, \tool demonstrates the necessary discernment to abstain when no valid location exists, returning a clear “no match found” response for over 70\% of the queries.

\begin{table*}[t]
    \centering
    \scriptsize 
    \setlength{\tabcolsep}{1.2pt} 
    \caption{Evaluation results on LogicQuery}
    \label{tab:comparison_lq}
    
    \begin{tabularx}{\textwidth}{@{} l l *{18}{C} @{}} 
        \toprule
        \multirow{2}{*}{\textbf{Methods}} & \multirow{2}{*}{\textbf{LLM}} & \multicolumn{6}{c}{\textbf{File Level (\%)}} & \multicolumn{6}{c}{\textbf{Module Level (\%)}} & \multicolumn{6}{c}{\textbf{Function Level (\%)}} \\
        \cmidrule(lr){3-8} \cmidrule(lr){9-14} \cmidrule(lr){15-20}
        & & SR & REC & PRE & AJS & PLR & HR & SR & REC & PRE & AJS & PLR & HR & SR & REC & PRE & AJS & PLR & HR \\
        
        \midrule
        \multirow{2}{*}{SweRank} 
        & Small & 8.00 & 7.40 & 11.13 & 6.64 & 0 & 31.11 & 4.17 & 3.74 & 5.24 & 3.45 & 0 & 24.07 & 1.02 & 1.87 & 1.28 & 0.74 & 0 & 11.73 \\
        & Large & 4.44 & 5.44 & 10.26 & 4.66 & 0 & 23.56 & 1.85 & 2.64 & 3.61 & 1.91 & 0 & 14.81 & 1.53 & 1.79 & 1.48 & 0.82 & 0 & 11.22 \\

        \midrule
        \multirow{2}{*}{Agentless} 
        & GPT-4o & 13.33 & 12.16 & 15.66 & 10.07 & 2.22 & 32.44 & 8.33 & 6.29 & 6.93 & 4.23 & 0 & 24.54 & 5.10 & 1.88 & 2.04 & 1.49 & 0 & 14.29 \\
        & Claude-3.5 & 12.00 & 15.14 & 18.99 & 11.46 & 2.67 & 34.22 & 6.02 & 7.62 & 9.05 & 5.71 & 0.46 & 24.54 & 4.08 & 3.53 & 4.87 & 3.29 & 0 & 15.31 \\

        \midrule
        \multirow{2}{*}{LocAgent} 
        & GPT-4o & 17.33 & 8.42 & 8.44 & 6.21 & 0.89 & 42.22 & 7.87 & 4.89 & 4.03 & 2.81 & 0 & 29.63 & 5.10 & 1.06 & 1.60 & 1.08 & 0 & 17.35 \\
        & Claude-3.5 & 22.22 & 11.30 & 11.02 & 8.73 & 0.44 & 54.67 & 15.74 & 7.76 & 6.63 & 4.97 & 0 & 47.69 & 11.73 & 5.81 & 4.32 & 3.22 & 0 & 40.82 \\

        \midrule
        \multirow{2}{*}{CoSIL} 
        & GPT-4o & 10.22 & 12.82 & 15.62 & 9.39 & 2.22 & 32.89 & 6.94 & 8.76 & 9.47 & 5.06 & 0 & 24.54 & 3.06 & 3.45 & 3.42 & 1.87 & 0 & 10.71 \\
        & Claude-3.5 & 14.22 & 13.60 & 16.16 & 10.39 & 1.33 & 34.67 & 10.19 & 9.04 & 10.17 & 6.50 & 0.93 & 27.78 & 6.12 & 5.42 & 5.87 & 3.57 & 0 & 19.39 \\

        \midrule
        \multirow{2}{*}{\makecell{\tool}} 
        & Claude-3.5 & 64.44 & 60.61 & 68.96 & 58.87 & 42.22 & \textbf{89.78} & 59.72 & 55.69 & 65.37 & 54.49 & 35.65 & \textbf{88.89} & 57.14 & 52.09 & 62.11 & 51.19 & 32.65 & \textbf{88.78} \\
        & Qwen3-Max & \textbf{68.44} & \textbf{68.43} & \textbf{73.35} & \textbf{65.28} & \textbf{48.44} & 87.11 & \textbf{62.96} & \textbf{63.87} & \textbf{70.40} & \textbf{61.44} & \textbf{42.59} & 86.11 & \textbf{58.67} & \textbf{61.65} & \textbf{66.98} & \textbf{58.03} & \textbf{38.27} & 84.69 \\
        
        \bottomrule
    \end{tabularx}
\end{table*}

% \begin{table}[htbp]
%     \centering
%     \caption{Summary of Model Performance on Null-Response Queries}
%     \label{tab:simplified_eval}
%     \begin{tabularx}{0.8\linewidth}{X c c c c c}
%         \toprule
%         \textbf{Methods} & \textbf{Total} & \textbf{TN} & \textbf{FP} & \textbf{Accuracy} & \textbf{Precision} \\
%         \midrule
%         Agentless & 25 & 25 & 0 & 100\% & 1.00 \\
%         LocAgent & 25 & 25 & 0 & 100\% & 1.00 \\
%         CoSIL & 25 & 25 & 0 & 100\% & 1.00 \\
%         Orca Loca & 25 & 25 & 0 & 100\% & 1.00 \\
%         DataLoc & 25 & 25 & 0 & 100\% & 1.00 \\
%         \bottomrule
%     \end{tabularx}
    
%     \vspace{10pt}
%     \small
%     \textit{Note: TN (True Negative) represents correct empty responses, while FP (False Positive) indicates incorrect non-empty responses.}
% \end{table}

\smallskip
\noindent\shadowbox{%
  \begin{minipage}{0.98\columnwidth}
    \textbf{Answer to RQ1:}
		The results clearly show that \tool effectively handles the keyword-agnostic logical code localization challenge, whereas baselines perform poorly when keyword shorts are unavailable. This means these approaches still rely on shallow lexical matching rather than genuine logical reasoning. Furthermore, their performance on the \negset exposes fundamental weakness in their refusal ability.
  \end{minipage}}
%%% Local Variables:
%%% mode: latex
%%% TeX-master: "../main"
%%% End:

\subsubsection{Effectiveness for general issues}

To evaluate the practical utility of \tool in real-world software maintenance, we conducted a comparative analysis on the SWE-bench Lite. Existing approaches usually employ top-$n$ strategy, whereas \tool operates without a predefined $n$. Table~\ref{tab:comparison_isq} illustrates that \tool remains highly competitive.

A critical distinction lies in the recommendation density and target precision. While \tool provides instance-specific localization with an average of only 2 candidates per issue, SOTA baselines rely on much broader and often fixed-size candidate sets to improve their accuracy. Specifically, CoSIL and LocAgent typically default to a top-$5$ recommendation at the function level, while Agentless routinely recommends 5 locations as candidates regardless of the issue's actual complexity. SweRank adopts an even more aggressive strategy, utilizing top-100 rankings.

Consequently, even when Acc@$n$ metrics appear comparable, \tool achieves substantially higher precision. By providing a concise and accurate set of entry points, \tool minimizes the noise that developers or downstream agents must filter, reducing validation overhead. This precision-centric design also yields substantial gains in resource efficiency (details in Section~\ref{sec:cost}). 

% Compared to agent-based baselines that require extensive multi-turn interactions, \tool operates with remarkably low latency and token consumption by focusing only on logically necessary code segments. We also observed that this efficiency contrasts with the operational instability of tools like LocAgent, which frequently falls into infinite execution loops on complex or mutated instances. While we will provide a comprehensive breakdown of these overhead metrics in RQ3, these preliminary results highlight that \tool delivers a superior balance of competitive recall, surgical precision, and minimal resource expenditure.

\begin{table}[t]
\centering
\caption{Comparison of different methods and models across various localization granularities.}
\label{tab:comparison_isq}
\small 
\begin{tabularx}{\textwidth}{@{} ll CCCCC CC @{}}
\toprule
\multirow{2}{*}{\textbf{Method}} & \multirow{2}{*}{\textbf{Model}} & \multicolumn{3}{c}{\textbf{File (\%)}} & \multicolumn{2}{c}{\textbf{Module (\%)}} & \multicolumn{2}{c}{\textbf{Function (\%)}} \\ 
\cmidrule(lr){3-5} \cmidrule(lr){6-7} \cmidrule(lr){8-9}
& & Acc@1 & Acc@3 & Acc@5 & Acc@5 & Acc@10 & Acc@5 & Acc@10 \\ 
\midrule
\multirow{2}{*}{SweRank}   & Small           & 78.10 & 92.34 & 94.53 & 89.05 & 92.70 & 79.56 & 86.13 \\
                           & Large           & 83.21 & 94.89 & 95.99 & 90.88 & 93.43 & 81.39 & 88.69 \\ 
\midrule
% \addlinespace[0.5em] 
\multirow{2}{*}{Agentless} & gpt-4o          & 67.50 & 74.45 & 74.45 & 67.15 & 67.15 & 55.47 & 55.47 \\
                           & claude-3.5      & 72.63 & 79.20 & 79.56 & 68.98 & 68.98 & 58.76 & 58.76 \\ 
\midrule
% \addlinespace[0.5em]
\multirow{2}{*}{LocAgent}  & Qwen2.5-32B(ft) & 75.91 & 90.51 & 92.70 & 85.77 & 87.23 & 71.90 & 77.01 \\
                           & claude-3.5      & 77.74 & 91.97 & 94.16 & 86.50 & 87.59 & 73.36 & 77.37 \\ 
\midrule
% \addlinespace[0.5em]
\multirow{2}{*}{\tool}   & gpt-5.1         & 71.53 & 77.38 & 78.47 & 70.80 & 72.26 & 63.14 & 64.96 \\
                           & claude-3.5      & 72.26 & 80.66 & 81.02 & 75.55 & 75.55 & 68.98 & 68.98 \\ 
\bottomrule
\end{tabularx}
\end{table}

\smallskip
\noindent\shadowbox{%
  \begin{minipage}{0.98\columnwidth}
    \textbf{Answer to RQ2:}
		\tool also demonstrates competitive performance on issue-solving benchmarks. Notably, given that \tool produces only 2 candidate locations on average, it offers distinct advantages in recommendation efficiency and in excluding incorrect locations. This precise localization helps reduce the potential overhead of downstream tasks.
  \end{minipage}}
\subsubsection{Efficiency}\label{sec:cost}
\begin{figure}[t]
	\centering
	\includegraphics[width=\textwidth]{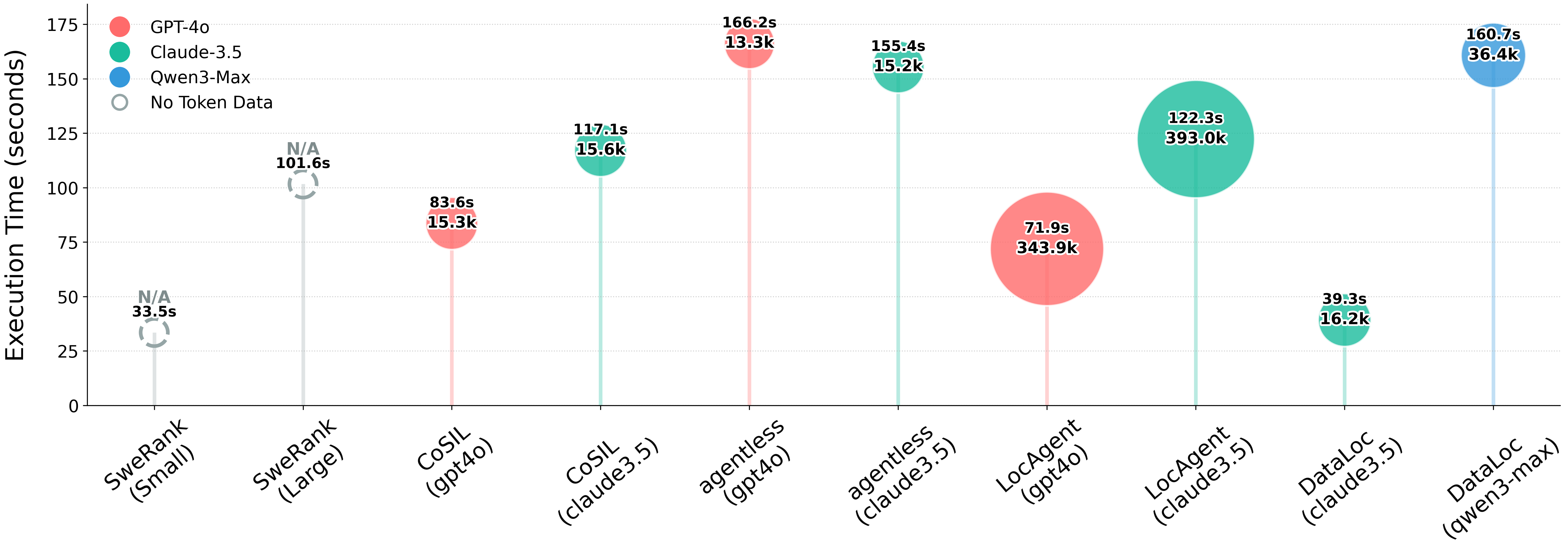}
	\caption{Average execution time and token consumption of \dataset}
	\label{fig:cost_analysis}
\end{figure}

Figure~\ref{fig:cost_analysis} presents a comprehensive comparison using a lollipop chart, where the vertical axis represents average execution time (seconds) and bubble size reflects total token consumption (labeled in thousands). As illustrated, \tool (Claude3.5) establishes a new efficiency frontier for KA-LCL tasks. It achieves a superior trade-off between temporal efficiency and resource consumption; while maintaining the shortest average execution time (39s), \tool also operates within the top-tier of average token consumption (16.2k).

\paragraph{Temporal efficiency.} \tool (Claude-3.5) completes localization in around half minute per task, outperforming all agentic baselines. Even the fastest LocAgent (GPT-4o) remains 2$\times$ slower. Crucially, the higher execution time of baselines does not yield better performance, as shown in Table~\ref{tab:comparison_lq}. This indicates that, without keyword shortcuts, baseline methods fail to perform meaningful repo-level inference, despite exhibiting intermediate reasoning steps. Additionally, we observe that, switching the model to Qwen3-Max significantly increases execution time. This occurs because Qwen3-Max frequently misuses Datalog syntax, which triggers our syntax correction, intermediate rule diagnostics, and mutation-based feedback more often. While these mechanisms ensure grammatical accuracy and improve localization precision, they inevitably introduce additional temporal overhead.

\paragraph{Token economy.} Approaches that rely more on agents incur substantially higher token consumption, with LocAgent and Agentless consuming over an order of magnitude more tokens per task than \tool. This token explosion reflects a trade-off where tokens are exchanged for intelligence, but this intelligence is currently tie to text. As a result, without keyword shortcuts to guide retrieval, these agents are trapped in multi-round conversation and iterative codebase exploration. In our evaluation, more than ten queries causes LocAgent to enter infinite loops without producing results.

The efficiency of \tool stems from its hybrid architecture. In our design, the LLM-based agent is strategically confined to high-level tasks: query analysis, Datalog program synthesis, and final candidate verification. By shifting deep inference to a specialized engine, \tool greatly reduces the overhead of redundant multi-round exploration. In addition, errors carry small cost, requiring only the regeneration of a Datalog program. Beyond this low overhead, our parser-gated validation and intermediate-rule feedback actively assist LLM to synthesize high-quality programs.

% Crucially, the computationally intensive processes of logical tracing and localization are offloaded to an offline efficient inference engine. By delegating deep reasoning to a specialized Datalog-based engine, \tool bypasses the redundant, multi-turn exploration cycles that typically plague pure LLM agents, where models must repeatedly guess or navigate through large codebases. Generating Datalog queries incurs minimal token overhead, and our syntax corrector ensures their correctness and usability. Even when errors occur, the token cost for query refinement remains low. This approach enables \tool to significantly reduce token consumption while maintaining accuracy, as the LLM no longer needs to process extensive raw code contexts or manage complex state space searches. Consequently, \tool transforms the localization problem from a high-cost generative search into an efficient, deterministic inference task, providing a scalable solution for the demanding KA-LCL challenge.

\smallskip
\noindent\shadowbox{%
  \begin{minipage}{0.98\columnwidth}
    \textbf{Answer to RQ3:}
		\tool defines the efficiency frontier in the KA-LCL challenge, maintaining an average execution time of 39.3s and a token consumption of 16.2k. By replacing expensive LLM-based exploration with Datalog-driven inference, \tool bypasses the prohibitive costs and execution loops of existing agentic baselines, demonstrating its potential industrial deployment.
  \end{minipage}}
\subsubsection{Ablation Study}\label{sec:ablation}
We design an ablation study to quantify how two mechanisms detailed in~\cref{sec:method:val} and \cref{sec:method:int} improve the quality of LLM-generated Datalog program.

%where ``quality'' encompasses (i) executability (passing parsing and evaluation), (ii) result usefulness (avoiding silent empty outputs caused by mistakes), and (iii) convergence efficiency (reaching a working query with fewer LLM iterations).

%For brevity, we refer to these two types of mechanism as \emph{validation and repair} (VAL) and \emph{intermediate-rule feedback} (INT), respectively.

\paragraph{Configurations}
We evaluate three configurations that progressively enable these mechanisms:
\textbf{Base} directly executes the LLM-generated Datalog program without any validation or additional feedback, except the output or error messages from \souffle{} itself. \textbf{VAL} (validation and repair) enables parser-gated validation with deterministic syntactic repairs and high-confidence semantic checks. \textbf{Full} further enables intermediate-rule mutation and feedback.
All configurations share the same LLM, initial prompt, iteration budget, and underlying fact bases.
%Each natural-language query is evaluated independently.

\paragraph{Metrics}
We compare (i) execution success rate and non-empty result rate, (ii) mean LLM iterations to first successful execution and (iii) final answer correctness. This evaluation is conducted on a representative subset of \dataset, consisting of the first 25 queries.

\begin{table}[t]
  \centering
  \small
  \setlength{\tabcolsep}{4.0pt}
  \renewcommand{\arraystretch}{1.15}
  \caption{Ablation results under two LLMs. \textbf{Base}: no mechanism;
  \textbf{VAL}: validation \& repair; \textbf{Full}: VAL+intermediate-rule feedback (INT).
  Rates are reported in \%. Iteration and time metrics report mean values (lower is better). Time is reported in seconds.}
  \label{tab:ablation}

  \resizebox{\columnwidth}{!}{%
  \begin{tabular}{ll
                  cc
                  cc c
                  cccccc}
    \toprule
    \multirow{2}{*}{\textbf{Model}} &
    \multirow{2}{*}{\textbf{Config}} &
    \multicolumn{2}{c}{\textbf{Rates} $\uparrow$} &
    \multicolumn{3}{c}{\textbf{Cost (Iter, Time)} $\downarrow$} &
    \multicolumn{6}{c}{\textbf{Correctness (function-level)} $\uparrow$} \\
    \cmidrule(lr){3-4}
    \cmidrule(lr){5-7}
    \cmidrule(lr){8-13}
    & & \textbf{ExecSucc} & \textbf{$\ne\varnothing$} &
    \textbf{First} & \textbf{Iter} & \textbf{Time} &
    \textbf{SR} & \textbf{PRE} & \textbf{REC} & \textbf{AJS} & \textbf{PLR} & \textbf{HR} \\
    \midrule

    \multirow{3}{*}{Qwen3-Max} &
    Base   & 73.59 & 49.41 & 1.48 & 12.92 & 427 & 20.00 & 18.20 & 15.87 & 26.73 & 16.00 & 20.00 \\
    & VAL  & 84.12 & 75.68 & 1.28 & 10.16 & 104 & 40.00 & 38.93 & 31.87 & 43.67 & 32.00 & 44.00 \\
    & Full & 84.12 & 78.52 & 1.24 & 10.92 & 136 & 56.00 & 55.20 & 45.53 & 58.73 & 40.00 & 60.00 \\
    \midrule

    \multirow{3}{*}{Claude-3.5} &
    Base   & 89.42 & 67.95 & 1.16 & 4.20 & 38 & 44.00 & 38.10 & 33.87 & 43.70 & 24.00 & 52.00 \\
    & VAL  & 94.67 & 75.86 & 1.04 & 4.28 & 40 & 44.00 & 38.90 & 31.87 & 48.36 & 32.00 & 52.00 \\
    & Full & 94.67 & 80.69 & 1.08 & 4.08 & 37 & 61.90 & 62.70 & 63.63 & 57.09 & 42.86 & 80.95 \\
    \bottomrule
  \end{tabular}}

  \vspace{1mm}
  \footnotesize\raggedright
  \textbf{Cost metrics.}
  \textbf{$\ne\varnothing$} denotes the fraction of queries that produce non-empty answers.
  \textbf{First} is the mean number of LLM iterations to the first successfully executing program.
  \textbf{Iter} is the average total number of LLM iterations consumed per query.
  \textbf{Time} reports the mean end-to-end wall-clock time per query, including failed attempts and tool feedback.
\end{table}

\Cref{tab:ablation} summarizes the ablation results under two different LLMs, comparing the baseline system with progressively enabled mechanisms.
The results show that validation and repair (VAL) substantially improves the quality of LLM-generated Datalog for both models, with a markedly stronger effect on Qwen3 Max than on Claude 3.5 Sonnet. This difference is expected given the models' baseline capabilities: without VAL, Qwen3 Max exhibits a significantly lower execution success rate due to its weaker ability to consistently produce syntactically correct \souffle{} Datalog, whereas Claude already achieves a relatively high baseline level of syntactic validity.

For Qwen3 Max, enabling VAL leads to a dramatic improvement across nearly all metrics. The non-empty result rate increases from 49\% to 75\%, indicating that a large fraction of previously failing or unproductive queries were recoverable once parser-gated validation and conservative repairs were applied. At the same time, the average end-to-end execution time drops sharply from 427 seconds to 104 seconds, reflecting a reduction in wasted iterations caused by unrecoverable parser errors and repeated failed executions. Improvements are also reflected in downstream task quality: function-level correctness metrics show substantial gains, with precision increasing from 18\% to 50\%, demonstrating that VAL does not merely enable execution but also materially improves the semantic adequacy of the resulting queries.
In contrast, the effect of VAL on Claude is more moderate: execution success rate increases by +5 percentage points, and the non-empty result rate increases by +8 percentage points. 

Enabling intermediate-rule feedback (INT) in the Full configuration produces a different pattern. The primary role of INT is to guide the LLM toward identifying which specific rule is semantically invalid in the sense of producing no results, and how that rule can be locally revised. As a result, INT may slightly increase iteration count or execution time in some cases due to additional diagnostic executions; however, this overhead consistently translates into higher non-empty rates and improved final answer correctness.

\smallskip
\noindent\shadowbox{%
  \begin{minipage}{0.98\columnwidth}
    \textbf{Answer to RQ4:} Validation and repair (VAL) substantially improves the robustness of
    LLM-generated Datalog, with particularly large gains for weaker models by increasing non-empty
    result rates and reducing execution cost. Intermediate-rule feedback (INT) complements VAL by
    guiding targeted revisions of logically unproductive rules, occasionally incurring additional
    diagnostic cost but improving final answer correctness across models.
  \end{minipage}}
%%% Local Variables:
%%% mode: latex
%%% TeX-master: "../main"
%%% End:

%%% Local Variables:
%%% mode: latex
%%% TeX-master: "../main"
%%% End:

\section{Threats to Validity}\label{sec:threats}
\paragraph{Internal Validity.} The primary internal threats concern baseline implementation and data leakage. We use the official implementations of all baselines with default configurations. For baselines that rely on large language models, we employ the same underlying models (GPT-4o and Claude-3.5-Sonnet) to avoid model-related bias. The KA-LCL queries in \dataset and \negset are constructed based on the decomposed code structure, resulting in novel query instances. Candidate ground truth are generated by state-of-the-art models (GPT-5.2, Claude-4.5-Opus, and Gemini-3-Pro) under advanced AIDE's agent mode and determined through independent manual verification by two authors. These measures eliminate the risk of data leakage.

\paragraph{External Validity.} A main threat to external validity is that our current evaluation focused only on Python. Although the proposed framework is language-agnostic in principle, extending it to additional languages or incorporating more analysis results into program facts remains future work, and addressing this challenge needs further engineering efforts to broaden the applicability. 
\section{Related Work}\label{sec:related}
\noindent \textbf{LLM for Issue Resolution and Question Answering.} 
Large language models are increasingly integrated into software engineering, motivating a wide range of benchmarks for codebase question answering and issue resolution. Existing benchmarks fall into two main categories. Code QA benchmarks~\cite{strich2024improving, li2024infibench, li2024procqa} evaluate models on either code snippets (e.g. CodeQueries~\cite{sahu2024codequeries}, CodeQA~\cite{liu2021codeqa}, CoSQA~\cite{liu2021codeqa}) or repository-level contexts derived from GitHub issues (e.g. CodeRepoQA~\cite{hu2024coderepoqa}, CoReQA~\cite{chen2025coreqa}, SWE-QA~\cite{peng2025swe}). End-to-end issue resolution benchmarks, such as SWE-Bench~\cite{jimenez2023swe} and its extensions, assess full issue-solving capabilities~\cite{jimenez2023swe,chen-etal-2025-locagent,zhuo2024bigcodebench,deng2025nocode,niu2023crosscodebench,chen2021evaluating,ouyang2024benchmarking,gao2023benchmarking,jiang2024collubenchbenchmarkpredictinglanguage,jain2024r2e,mundler2024swt,xie2024osworld, yang2025swesmith}, primarily focus on bug-fix tasks and limited programming languages. However, excessive keywords in these benchmarks provide models with too many shortcuts for code localization by superficial lexical matching. To mitigate this bias, we propose \dataset, which removes semantic keywords and only keeps logic structures, and \negset, an empty-ground-truth variant designed to evaluate abstention ability.

\noindent \textbf{Code Localization.}  
Code localization refers to identifying relevant code locations (e.g., files, modules, or functions) to resolve developers' queries. 
Recent advancements in this area have taken two complementary directions. 
Meanwhile, LLM-based retrieval techniques have been proposed to improve code localization performance by leveraging semantic understanding~\cite{xia2024agentless,chen-etal-2025-locagent,reddy2025swerank,wang2024openhands,yang2024swe,jiang2025cosil,zhang2024autocoderover,tao2024magis,xie2025swe,ma2025sorft}.
Recent research has proposed numerous code localization approaches that can be broadly categorized
into three classes: (1) \textit{Embedding-based approaches} (e.g.,
SWERankEmbed~\cite{reddy2025swerank}, CodeSage~\cite{zhang2024code}) encode code entities and natural language descriptions as
embeddings, ranking them based on semantic similarity. While they achieve high recall by retrieving
a broad range of relevant candidate locations, they can only identify code snippets that ``look
similar'' without understanding the logical relationships between them.
Furthermore, they suffer from hallucination and noise interference, such as methods with identical
names but entirely different functionalities.
(2) \textit{Pipeline-based LLM approaches} (e.g., Agentless~\cite{xia2024agentless}) follow a
structured, multi-stage workflow from files to functions.
However, such hierarchical localization design overly depends on initial file-level
localization and fails to capture cross-level dependencies, implicitly assuming that developers
adhere to good naming conventions.
(3) \textit{Agent-based LLM approches} (e.g., LocAgent~\cite{chen-etal-2025-locagent},
CoSIL~\cite{jiang2025cosil}, Orca Loca~\cite{yu2025orcalocallmagentframework}, GraphLocator~\cite{liu2025graphlocator}) offer greater flexibility by
allowing LLMs to autonomously traverse the repository graph.
Nevertheless, they only consider surface-level relevance and exhibit rapid performance degradation
without explicit contextual guidance.

% AIDE, such as \texttt{Cursor}~\cite{cursor_ai_editor}, \texttt{Gemini-CLI}~\cite{gemini_cli}, and \texttt{Claude-code}~\cite{claude_code}, have emerged to assist developers by suggesting or navigating to relevant code. 

% CodeQA
% CodeQL
% code search
% BM25 retrieval

\section{Conclusion}\label{sec:conclusion}
In this paper, we critically examined the limitations of contemporary code localization paradigms, uncovering a pervasive reliance on keyword shortcuts. By introducing a novel diagnostic benchmark, \dataset, we demonstrated that existing solutions rely excessively on the model's sophisticated lexical matching capabilities, thus struggling with keyword-agnostic logical code localization challenge. Another \negset, which deliberately leaves ground truth empty, reveals the trade-off between top-$n$ recommendations and false positives. To overcome these hurdles, we proposed \tool, a neuro-symbolic framework that integrates the semantic power of LLMs with the formal rigor of Datalog. \tool transform natural language intent into deterministic logic queries and apply mutations on intermediate rules to guarantee the quality of queries in advance, thus providing verifiable localization results. Our extensive evaluation confirms that \tool not only bridges the reasoning gap in complex logical queries but also offers a more resource-efficient and reliable solution for autonomous software engineering. This is a meaningful attempt at neuro-symbolic approaches, suggesting that formal methods can offer a path toward determinism in AI systems.

\section{Data Availability}
The source code and dataset are available at \url{https://anonymous.4open.science/r/LogicLoc-D9BE}

\bibliographystyle{ACM-Reference-Format}
\bibliography{acmart}


\begin{thebibliography}{37}

%%% the Web address.

\ifx \showCODEN    \undefined \def \showCODEN     #1{\unskip}     \fi
\ifx \showDOI      \undefined \def \showDOI       #1{#1}\fi
\ifx \showISBNx    \undefined \def \showISBNx     #1{\unskip}     \fi
\ifx \showISBNxiii \undefined \def \showISBNxiii  #1{\unskip}     \fi
\ifx \showISSN     \undefined \def \showISSN      #1{\unskip}     \fi
\ifx \showLCCN     \undefined \def \showLCCN      #1{\unskip}     \fi
\ifx \shownote     \undefined \def \shownote      #1{#1}          \fi
\ifx \showarticletitle \undefined \def \showarticletitle #1{#1}   \fi
\ifx \showURL      \undefined \def \showURL       {\relax}        \fi
% invisible to TeX
\providecommand\bibfield[2]{#2}
\providecommand\bibinfo[2]{#2}
\providecommand\natexlab[1]{#1}
\providecommand\showeprint[2][]{arXiv:#2}

\bibitem[Chen et~al\mbox{.}(2025b)]%
        {chen2025coreqa}
\bibfield{author}{\bibinfo{person}{Jialiang Chen}, \bibinfo{person}{Kaifa
  Zhao}, \bibinfo{person}{Jie Liu}, \bibinfo{person}{Chao Peng},
  \bibinfo{person}{Jierui Liu}, \bibinfo{person}{Hang Zhu},
  \bibinfo{person}{Pengfei Gao}, \bibinfo{person}{Ping Yang}, {and}
  \bibinfo{person}{Shuiguang Deng}.} \bibinfo{year}{2025}\natexlab{b}.
\newblock \showarticletitle{CoreQA: uncovering potentials of language models in
  code repository question answering}.
\newblock \bibinfo{journal}{\emph{arXiv preprint arXiv:2501.03447}}
  (\bibinfo{year}{2025}).
\newblock


\bibitem[Chen et~al\mbox{.}(2021)]%
        {chen2021evaluating}
\bibfield{author}{\bibinfo{person}{Mark Chen}, \bibinfo{person}{Jerry Tworek},
  \bibinfo{person}{Heewoo Jun}, \bibinfo{person}{Qiming Yuan},
  \bibinfo{person}{Henrique Ponde De~Oliveira Pinto}, \bibinfo{person}{Jared
  Kaplan}, \bibinfo{person}{Harri Edwards}, \bibinfo{person}{Yuri Burda},
  \bibinfo{person}{Nicholas Joseph}, \bibinfo{person}{Greg Brockman},
  {et~al\mbox{.}}} \bibinfo{year}{2021}\natexlab{}.
\newblock \showarticletitle{Evaluating large language models trained on code}.
\newblock \bibinfo{journal}{\emph{arXiv preprint arXiv:2107.03374}}
  (\bibinfo{year}{2021}).
\newblock


\bibitem[Chen et~al\mbox{.}(2025a)]%
        {chen-etal-2025-locagent}
\bibfield{author}{\bibinfo{person}{Zhaoling Chen}, \bibinfo{person}{Robert
  Tang}, \bibinfo{person}{Gangda Deng}, \bibinfo{person}{Fang Wu},
  \bibinfo{person}{Jialong Wu}, \bibinfo{person}{Zhiwei Jiang},
  \bibinfo{person}{Viktor Prasanna}, \bibinfo{person}{Arman Cohan}, {and}
  \bibinfo{person}{Xingyao Wang}.} \bibinfo{year}{2025}\natexlab{a}.
\newblock \showarticletitle{{L}oc{A}gent: Graph-Guided {LLM} Agents for Code
  Localization}. In \bibinfo{booktitle}{\emph{Proceedings of the 63rd Annual
  Meeting of the Association for Computational Linguistics (Volume 1: Long
  Papers)}}, \bibfield{editor}{\bibinfo{person}{Wanxiang Che},
  \bibinfo{person}{Joyce Nabende}, \bibinfo{person}{Ekaterina Shutova}, {and}
  \bibinfo{person}{Mohammad~Taher Pilehvar}} (Eds.).
  \bibinfo{publisher}{Association for Computational Linguistics},
  \bibinfo{address}{Vienna, Austria}, \bibinfo{pages}{8697--8727}.
\newblock
\showISBNx{979-8-89176-251-0}
\urldef\tempurl%
\url{https://doi.org/10.18653/v1/2025.acl-long.426}
\showDOI{\tempurl}


\bibitem[Deng et~al\mbox{.}(2025)]%
        {deng2025nocode}
\bibfield{author}{\bibinfo{person}{Le Deng}, \bibinfo{person}{Zhonghao Jiang},
  \bibinfo{person}{Jialun Cao}, \bibinfo{person}{Michael Pradel}, {and}
  \bibinfo{person}{Zhongxin Liu}.} \bibinfo{year}{2025}\natexlab{}.
\newblock \showarticletitle{NoCode-bench: A Benchmark for Evaluating Natural
  Language-Driven Feature Addition}.
\newblock \bibinfo{journal}{\emph{arXiv preprint arXiv:2507.18130}}
  (\bibinfo{year}{2025}).
\newblock


\bibitem[Di~Grazia and Pradel(2023)]%
        {di2023code}
\bibfield{author}{\bibinfo{person}{Luca Di~Grazia} {and}
  \bibinfo{person}{Michael Pradel}.} \bibinfo{year}{2023}\natexlab{}.
\newblock \showarticletitle{Code search: A survey of techniques for finding
  code}.
\newblock \bibinfo{journal}{\emph{Comput. Surveys}} \bibinfo{volume}{55},
  \bibinfo{number}{11} (\bibinfo{year}{2023}), \bibinfo{pages}{1--31}.
\newblock


\bibitem[Gao et~al\mbox{.}(2023)]%
        {gao2023benchmarking}
\bibfield{author}{\bibinfo{person}{Xinyu Gao}, \bibinfo{person}{Zhijie Wang},
  \bibinfo{person}{Yang Feng}, \bibinfo{person}{Lei Ma},
  \bibinfo{person}{Zhenyu Chen}, {and} \bibinfo{person}{Baowen Xu}.}
  \bibinfo{year}{2023}\natexlab{}.
\newblock \showarticletitle{Benchmarking robustness of ai-enabled multi-sensor
  fusion systems: Challenges and opportunities}. In
  \bibinfo{booktitle}{\emph{Proceedings of the 31st ACM Joint European Software
  Engineering Conference and Symposium on the Foundations of Software
  Engineering}}. \bibinfo{pages}{871--882}.
\newblock


\bibitem[Hu et~al\mbox{.}(2024)]%
        {hu2024coderepoqa}
\bibfield{author}{\bibinfo{person}{Ruida Hu}, \bibinfo{person}{Chao Peng},
  \bibinfo{person}{Jingyi Ren}, \bibinfo{person}{Bo Jiang},
  \bibinfo{person}{Xiangxin Meng}, \bibinfo{person}{Qinyun Wu},
  \bibinfo{person}{Pengfei Gao}, \bibinfo{person}{Xinchen Wang}, {and}
  \bibinfo{person}{Cuiyun Gao}.} \bibinfo{year}{2024}\natexlab{}.
\newblock \showarticletitle{CodeRepoQA: A Large-scale Benchmark for Software
  Engineering Question Answering}.
\newblock \bibinfo{journal}{\emph{arXiv preprint arXiv:2412.14764}}
  (\bibinfo{year}{2024}).
\newblock


\bibitem[Jain et~al\mbox{.}(2024)]%
        {jain2024r2e}
\bibfield{author}{\bibinfo{person}{Naman Jain}, \bibinfo{person}{Manish
  Shetty}, \bibinfo{person}{Tianjun Zhang}, \bibinfo{person}{King Han},
  \bibinfo{person}{Koushik Sen}, {and} \bibinfo{person}{Ion Stoica}.}
  \bibinfo{year}{2024}\natexlab{}.
\newblock \showarticletitle{R2e: Turning any github repository into a
  programming agent environment}. In \bibinfo{booktitle}{\emph{Forty-first
  International Conference on Machine Learning}}.
\newblock


\bibitem[Jiang et~al\mbox{.}(2024)]%
        {jiang2024collubenchbenchmarkpredictinglanguage}
\bibfield{author}{\bibinfo{person}{Nan Jiang}, \bibinfo{person}{Qi Li},
  \bibinfo{person}{Lin Tan}, {and} \bibinfo{person}{Tianyi Zhang}.}
  \bibinfo{year}{2024}\natexlab{}.
\newblock \bibinfo{title}{Collu-Bench: A Benchmark for Predicting Language
  Model Hallucinations in Code}.
\newblock
\newblock
\showeprint[arxiv]{2410.09997}~[cs.SE]
\urldef\tempurl%
\url{https://arxiv.org/abs/2410.09997}
\showURL{%
\tempurl}


\bibitem[Jiang et~al\mbox{.}(2025)]%
        {jiang2025cosil}
\bibfield{author}{\bibinfo{person}{Zhonghao Jiang}, \bibinfo{person}{Xiaoxue
  Ren}, \bibinfo{person}{Meng Yan}, \bibinfo{person}{Wei Jiang},
  \bibinfo{person}{Yong Li}, {and} \bibinfo{person}{Zhongxin Liu}.}
  \bibinfo{year}{2025}\natexlab{}.
\newblock \showarticletitle{CoSIL: Software Issue Localization via LLM-Driven
  Code Repository Graph Searching}.
\newblock \bibinfo{journal}{\emph{arXiv preprint arXiv:2503.22424}}
  (\bibinfo{year}{2025}).
\newblock


\bibitem[Jimenez et~al\mbox{.}(2024)]%
        {swebench2024}
\bibfield{author}{\bibinfo{person}{Carlos~E. Jimenez}, \bibinfo{person}{John
  Yang}, \bibinfo{person}{Alexander Wettig}, \bibinfo{person}{Shunyu Yao},
  \bibinfo{person}{Kexin Pei}, \bibinfo{person}{Ofir Press}, {and}
  \bibinfo{person}{Karthik Narasimhan}.} \bibinfo{year}{2024}\natexlab{}.
\newblock \bibinfo{title}{{SWE}-bench: Can Language Models Resolve Real-world
  Github Issues?}
\newblock
  \bibinfo{howpublished}{\url{https://huggingface.co/datasets/SWE-bench/SWE-bench_Lite}}.
\newblock
\newblock
\shownote{Accessed: 2025-10-04}.


\bibitem[Jimenez et~al\mbox{.}({[n.\,d.]})]%
        {jimenez2023swe}
\bibfield{author}{\bibinfo{person}{Carlos~E Jimenez}, \bibinfo{person}{John
  Yang}, \bibinfo{person}{Alexander Wettig}, \bibinfo{person}{Shunyu Yao},
  \bibinfo{person}{Kexin Pei}, \bibinfo{person}{Ofir Press}, {and}
  \bibinfo{person}{Karthik~R Narasimhan}.} \bibinfo{year}{[n.\,d.]}\natexlab{}.
\newblock \showarticletitle{SWE-bench: Can Language Models Resolve Real-world
  Github Issues?}. In \bibinfo{booktitle}{\emph{The Twelfth International
  Conference on Learning Representations}}.
\newblock


\bibitem[Li et~al\mbox{.}(2024a)]%
        {li2024infibench}
\bibfield{author}{\bibinfo{person}{Linyi Li}, \bibinfo{person}{Shijie Geng},
  \bibinfo{person}{Zhenwen Li}, \bibinfo{person}{Yibo He}, \bibinfo{person}{Hao
  Yu}, \bibinfo{person}{Ziyue Hua}, \bibinfo{person}{Guanghan Ning},
  \bibinfo{person}{Siwei Wang}, \bibinfo{person}{Tao Xie}, {and}
  \bibinfo{person}{Hongxia Yang}.} \bibinfo{year}{2024}\natexlab{a}.
\newblock \showarticletitle{Infibench: Evaluating the question-answering
  capabilities of code large language models}.
\newblock \bibinfo{journal}{\emph{Advances in Neural Information Processing
  Systems}}  \bibinfo{volume}{37} (\bibinfo{year}{2024}),
  \bibinfo{pages}{128668--128698}.
\newblock


\bibitem[Li et~al\mbox{.}(2024b)]%
        {li2024procqa}
\bibfield{author}{\bibinfo{person}{Zehan Li}, \bibinfo{person}{Jianfei Zhang},
  \bibinfo{person}{Chuantao Yin}, \bibinfo{person}{Yuanxin Ouyang}, {and}
  \bibinfo{person}{Wenge Rong}.} \bibinfo{year}{2024}\natexlab{b}.
\newblock \showarticletitle{ProCQA: a large-scale community-based programming
  question answering dataset for code search}.
\newblock \bibinfo{journal}{\emph{arXiv preprint arXiv:2403.16702}}
  (\bibinfo{year}{2024}).
\newblock


\bibitem[Liu and Wan(2021)]%
        {liu2021codeqa}
\bibfield{author}{\bibinfo{person}{Chenxiao Liu} {and} \bibinfo{person}{Xiaojun
  Wan}.} \bibinfo{year}{2021}\natexlab{}.
\newblock \showarticletitle{CodeQA: A question answering dataset for source
  code comprehension}.
\newblock \bibinfo{journal}{\emph{arXiv preprint arXiv:2109.08365}}
  (\bibinfo{year}{2021}).
\newblock


\bibitem[Liu et~al\mbox{.}(2025)]%
        {liu2025graphlocator}
\bibfield{author}{\bibinfo{person}{Wei Liu}, \bibinfo{person}{Chao Peng},
  \bibinfo{person}{Pengfei Gao}, \bibinfo{person}{Aofan Liu},
  \bibinfo{person}{Wei Zhang}, \bibinfo{person}{Haiyan Zhao}, {and}
  \bibinfo{person}{Zhi Jin}.} \bibinfo{year}{2025}\natexlab{}.
\newblock \showarticletitle{GraphLocator: Graph-guided Causal Reasoning for
  Issue Localization}.
\newblock \bibinfo{journal}{\emph{arXiv preprint arXiv:2512.22469}}
  (\bibinfo{year}{2025}).
\newblock


\bibitem[Ma et~al\mbox{.}(2025)]%
        {ma2025sorft}
\bibfield{author}{\bibinfo{person}{Zexiong Ma}, \bibinfo{person}{Chao Peng},
  \bibinfo{person}{Pengfei Gao}, \bibinfo{person}{Xiangxin Meng},
  \bibinfo{person}{Yanzhen Zou}, {and} \bibinfo{person}{Bing Xie}.}
  \bibinfo{year}{2025}\natexlab{}.
\newblock \showarticletitle{SoRFT: Issue Resolving with Subtask-oriented
  Reinforced Fine-Tuning}.
\newblock \bibinfo{journal}{\emph{CoRR}} (\bibinfo{year}{2025}).
\newblock


\bibitem[M{\"u}ndler et~al\mbox{.}(2024)]%
        {mundler2024swt}
\bibfield{author}{\bibinfo{person}{Niels M{\"u}ndler}, \bibinfo{person}{Mark
  M{\"u}ller}, \bibinfo{person}{Jingxuan He}, {and} \bibinfo{person}{Martin
  Vechev}.} \bibinfo{year}{2024}\natexlab{}.
\newblock \showarticletitle{SWT-bench: Testing and validating real-world
  bug-fixes with code agents}.
\newblock \bibinfo{journal}{\emph{Advances in Neural Information Processing
  Systems}}  \bibinfo{volume}{37} (\bibinfo{year}{2024}),
  \bibinfo{pages}{81857--81887}.
\newblock


\bibitem[Niu et~al\mbox{.}(2023)]%
        {niu2023crosscodebench}
\bibfield{author}{\bibinfo{person}{Changan Niu}, \bibinfo{person}{Chuanyi Li},
  \bibinfo{person}{Vincent Ng}, {and} \bibinfo{person}{Bin Luo}.}
  \bibinfo{year}{2023}\natexlab{}.
\newblock \showarticletitle{Crosscodebench: Benchmarking cross-task
  generalization of source code models}. In \bibinfo{booktitle}{\emph{2023
  IEEE/ACM 45th International Conference on Software Engineering (ICSE)}}.
  IEEE, \bibinfo{pages}{537--549}.
\newblock


\bibitem[Ouyang et~al\mbox{.}(2024)]%
        {ouyang2024benchmarking}
\bibfield{author}{\bibinfo{person}{Yicheng Ouyang}, \bibinfo{person}{Jun Yang},
  {and} \bibinfo{person}{Lingming Zhang}.} \bibinfo{year}{2024}\natexlab{}.
\newblock \showarticletitle{Benchmarking automated program repair: An extensive
  study on both real-world and artificial bugs}. In
  \bibinfo{booktitle}{\emph{Proceedings of the 33rd ACM SIGSOFT International
  Symposium on Software Testing and Analysis}}. \bibinfo{pages}{440--452}.
\newblock


\bibitem[Peng et~al\mbox{.}(2025)]%
        {peng2025swe}
\bibfield{author}{\bibinfo{person}{Weihan Peng}, \bibinfo{person}{Yuling Shi},
  \bibinfo{person}{Yuhang Wang}, \bibinfo{person}{Xinyun Zhang},
  \bibinfo{person}{Beijun Shen}, {and} \bibinfo{person}{Xiaodong Gu}.}
  \bibinfo{year}{2025}\natexlab{}.
\newblock \showarticletitle{SWE-QA: Can Language Models Answer Repository-level
  Code Questions?}
\newblock \bibinfo{journal}{\emph{arXiv preprint arXiv:2509.14635}}
  (\bibinfo{year}{2025}).
\newblock


\bibitem[Reddy et~al\mbox{.}(2025)]%
        {reddy2025swerank}
\bibfield{author}{\bibinfo{person}{Revanth~Gangi Reddy}, \bibinfo{person}{Tarun
  Suresh}, \bibinfo{person}{JaeHyeok Doo}, \bibinfo{person}{Ye Liu},
  \bibinfo{person}{Xuan~Phi Nguyen}, \bibinfo{person}{Yingbo Zhou},
  \bibinfo{person}{Semih Yavuz}, \bibinfo{person}{Caiming Xiong},
  \bibinfo{person}{Heng Ji}, {and} \bibinfo{person}{Shafiq Joty}.}
  \bibinfo{year}{2025}\natexlab{}.
\newblock \showarticletitle{SweRank: Software Issue Localization with Code
  Ranking}.
\newblock \bibinfo{journal}{\emph{arXiv preprint arXiv:2505.07849}}
  (\bibinfo{year}{2025}).
\newblock


\bibitem[Sahu et~al\mbox{.}(2024)]%
        {sahu2024codequeries}
\bibfield{author}{\bibinfo{person}{Surya~Prakash Sahu},
  \bibinfo{person}{Madhurima Mandal}, \bibinfo{person}{Shikhar Bharadwaj},
  \bibinfo{person}{Aditya Kanade}, \bibinfo{person}{Petros Maniatis}, {and}
  \bibinfo{person}{Shirish Shevade}.} \bibinfo{year}{2024}\natexlab{}.
\newblock \showarticletitle{Codequeries: A dataset of semantic queries over
  code}. In \bibinfo{booktitle}{\emph{Proceedings of the 17th Innovations in
  Software Engineering Conference}}. \bibinfo{pages}{1--11}.
\newblock


\bibitem[Strich et~al\mbox{.}(2024)]%
        {strich2024improving}
\bibfield{author}{\bibinfo{person}{Jan Strich}, \bibinfo{person}{Florian
  Schneider}, \bibinfo{person}{Irina Nikishina}, {and} \bibinfo{person}{Chris
  Biemann}.} \bibinfo{year}{2024}\natexlab{}.
\newblock \showarticletitle{On Improving Repository-Level Code QA for Large
  Language Models}. In \bibinfo{booktitle}{\emph{Proceedings of the 62nd Annual
  Meeting of the Association for Computational Linguistics (Volume 4: Student
  Research Workshop)}}. \bibinfo{pages}{209--244}.
\newblock


\bibitem[Suresh et~al\mbox{.}(2024)]%
        {suresh2024cornstack}
\bibfield{author}{\bibinfo{person}{Tarun Suresh}, \bibinfo{person}{Revanth
  Gangi~Reddy}, \bibinfo{person}{Yifei Xu}, \bibinfo{person}{Zach Nussbaum},
  \bibinfo{person}{Andriy Mulyar}, \bibinfo{person}{Brandon Duderstadt}, {and}
  \bibinfo{person}{Heng Ji}.} \bibinfo{year}{2024}\natexlab{}.
\newblock \showarticletitle{Cornstack: High-quality contrastive data for better
  code ranking}.
\newblock \bibinfo{journal}{\emph{arXiv e-prints}} (\bibinfo{year}{2024}),
  \bibinfo{pages}{arXiv--2412}.
\newblock


\bibitem[Tao et~al\mbox{.}(2024)]%
        {tao2024magis}
\bibfield{author}{\bibinfo{person}{Wei Tao}, \bibinfo{person}{Yucheng Zhou},
  \bibinfo{person}{Yanlin Wang}, \bibinfo{person}{Wenqiang Zhang},
  \bibinfo{person}{Hongyu Zhang}, {and} \bibinfo{person}{Yu Cheng}.}
  \bibinfo{year}{2024}\natexlab{}.
\newblock \showarticletitle{Magis: Llm-based multi-agent framework for github
  issue resolution}.
\newblock \bibinfo{journal}{\emph{Advances in Neural Information Processing
  Systems}}  \bibinfo{volume}{37} (\bibinfo{year}{2024}),
  \bibinfo{pages}{51963--51993}.
\newblock


\bibitem[Wang et~al\mbox{.}(2024)]%
        {wang2024openhands}
\bibfield{author}{\bibinfo{person}{Xingyao Wang}, \bibinfo{person}{Boxuan Li},
  \bibinfo{person}{Yufan Song}, \bibinfo{person}{Frank~F Xu},
  \bibinfo{person}{Xiangru Tang}, \bibinfo{person}{Mingchen Zhuge},
  \bibinfo{person}{Jiayi Pan}, \bibinfo{person}{Yueqi Song},
  \bibinfo{person}{Bowen Li}, \bibinfo{person}{Jaskirat Singh},
  {et~al\mbox{.}}} \bibinfo{year}{2024}\natexlab{}.
\newblock \showarticletitle{Openhands: An open platform for ai software
  developers as generalist agents}.
\newblock \bibinfo{journal}{\emph{arXiv preprint arXiv:2407.16741}}
  (\bibinfo{year}{2024}).
\newblock


\bibitem[Wu et~al\mbox{.}(2021)]%
        {wu2021diffbase}
\bibfield{author}{\bibinfo{person}{Xiuheng Wu}, \bibinfo{person}{Chenguang
  Zhu}, {and} \bibinfo{person}{Yi Li}.} \bibinfo{year}{2021}\natexlab{}.
\newblock \showarticletitle{Diffbase: A differential factbase for effective
  software evolution management}. In \bibinfo{booktitle}{\emph{Proceedings of
  the 29th ACM Joint Meeting on European Software Engineering Conference and
  Symposium on the Foundations of Software Engineering}}.
  \bibinfo{pages}{503--515}.
\newblock


\bibitem[Xia et~al\mbox{.}(2024)]%
        {xia2024agentless}
\bibfield{author}{\bibinfo{person}{Chunqiu~Steven Xia}, \bibinfo{person}{Yinlin
  Deng}, \bibinfo{person}{Soren Dunn}, {and} \bibinfo{person}{Lingming Zhang}.}
  \bibinfo{year}{2024}\natexlab{}.
\newblock \showarticletitle{Agentless: Demystifying llm-based software
  engineering agents}.
\newblock \bibinfo{journal}{\emph{arXiv preprint arXiv:2407.01489}}
  (\bibinfo{year}{2024}).
\newblock


\bibitem[Xie et~al\mbox{.}(2025)]%
        {xie2025swe}
\bibfield{author}{\bibinfo{person}{Chengxing Xie}, \bibinfo{person}{Bowen Li},
  \bibinfo{person}{Chang Gao}, \bibinfo{person}{He Du}, \bibinfo{person}{Wai
  Lam}, \bibinfo{person}{Difan Zou}, {and} \bibinfo{person}{Kai Chen}.}
  \bibinfo{year}{2025}\natexlab{}.
\newblock \showarticletitle{SWE-Fixer: Training Open-Source LLMs for Effective
  and Efficient GitHub Issue Resolution}. In \bibinfo{booktitle}{\emph{ICLR
  2025 Third Workshop on Deep Learning for Code}}.
\newblock


\bibitem[Xie et~al\mbox{.}(2024)]%
        {xie2024osworld}
\bibfield{author}{\bibinfo{person}{Tianbao Xie}, \bibinfo{person}{Danyang
  Zhang}, \bibinfo{person}{Jixuan Chen}, \bibinfo{person}{Xiaochuan Li},
  \bibinfo{person}{Siheng Zhao}, \bibinfo{person}{Ruisheng Cao},
  \bibinfo{person}{Toh~J Hua}, \bibinfo{person}{Zhoujun Cheng},
  \bibinfo{person}{Dongchan Shin}, \bibinfo{person}{Fangyu Lei},
  {et~al\mbox{.}}} \bibinfo{year}{2024}\natexlab{}.
\newblock \showarticletitle{Osworld: Benchmarking multimodal agents for
  open-ended tasks in real computer environments}.
\newblock \bibinfo{journal}{\emph{Advances in Neural Information Processing
  Systems}}  \bibinfo{volume}{37} (\bibinfo{year}{2024}),
  \bibinfo{pages}{52040--52094}.
\newblock


\bibitem[Yang et~al\mbox{.}(2024)]%
        {yang2024swe}
\bibfield{author}{\bibinfo{person}{John Yang}, \bibinfo{person}{Carlos~E
  Jimenez}, \bibinfo{person}{Alexander Wettig}, \bibinfo{person}{Kilian
  Lieret}, \bibinfo{person}{Shunyu Yao}, \bibinfo{person}{Karthik Narasimhan},
  {and} \bibinfo{person}{Ofir Press}.} \bibinfo{year}{2024}\natexlab{}.
\newblock \showarticletitle{Swe-agent: Agent-computer interfaces enable
  automated software engineering}.
\newblock \bibinfo{journal}{\emph{Advances in Neural Information Processing
  Systems}}  \bibinfo{volume}{37} (\bibinfo{year}{2024}),
  \bibinfo{pages}{50528--50652}.
\newblock


\bibitem[Yang et~al\mbox{.}(2025)]%
        {yang2025swesmith}
\bibfield{author}{\bibinfo{person}{John Yang}, \bibinfo{person}{Kilian Lieret},
  \bibinfo{person}{Carlos~E. Jimenez}, \bibinfo{person}{Alexander Wettig},
  \bibinfo{person}{Kabir Khandpur}, \bibinfo{person}{Yanzhe Zhang},
  \bibinfo{person}{Binyuan Hui}, \bibinfo{person}{Ofir Press},
  \bibinfo{person}{Ludwig Schmidt}, {and} \bibinfo{person}{Diyi Yang}.}
  \bibinfo{year}{2025}\natexlab{}.
\newblock \bibinfo{title}{SWE-smith: Scaling Data for Software Engineering
  Agents}.
\newblock
\newblock
\showeprint[arxiv]{2504.21798}~[cs.SE]
\urldef\tempurl%
\url{https://arxiv.org/abs/2504.21798}
\showURL{%
\tempurl}


\bibitem[Yu et~al\mbox{.}(2025)]%
        {yu2025orcalocallmagentframework}
\bibfield{author}{\bibinfo{person}{Zhongming Yu}, \bibinfo{person}{Hejia
  Zhang}, \bibinfo{person}{Yujie Zhao}, \bibinfo{person}{Hanxian Huang},
  \bibinfo{person}{Matrix Yao}, \bibinfo{person}{Ke Ding}, {and}
  \bibinfo{person}{Jishen Zhao}.} \bibinfo{year}{2025}\natexlab{}.
\newblock \bibinfo{title}{OrcaLoca: An LLM Agent Framework for Software Issue
  Localization}.
\newblock
\newblock
\showeprint[arxiv]{2502.00350}~[cs.SE]
\urldef\tempurl%
\url{https://arxiv.org/abs/2502.00350}
\showURL{%
\tempurl}


\bibitem[Zhang et~al\mbox{.}(2024a)]%
        {zhang2024code}
\bibfield{author}{\bibinfo{person}{Dejiao Zhang}, \bibinfo{person}{Wasi Ahmad},
  \bibinfo{person}{Ming Tan}, \bibinfo{person}{Hantian Ding},
  \bibinfo{person}{Ramesh Nallapati}, \bibinfo{person}{Dan Roth},
  \bibinfo{person}{Xiaofei Ma}, {and} \bibinfo{person}{Bing Xiang}.}
  \bibinfo{year}{2024}\natexlab{a}.
\newblock \showarticletitle{Code representation learning at scale}.
\newblock \bibinfo{journal}{\emph{arXiv preprint arXiv:2402.01935}}
  (\bibinfo{year}{2024}).
\newblock


\bibitem[Zhang et~al\mbox{.}(2024b)]%
        {zhang2024autocoderover}
\bibfield{author}{\bibinfo{person}{Yuntong Zhang}, \bibinfo{person}{Haifeng
  Ruan}, \bibinfo{person}{Zhiyu Fan}, {and} \bibinfo{person}{Abhik
  Roychoudhury}.} \bibinfo{year}{2024}\natexlab{b}.
\newblock \showarticletitle{Autocoderover: Autonomous program improvement}. In
  \bibinfo{booktitle}{\emph{Proceedings of the 33rd ACM SIGSOFT International
  Symposium on Software Testing and Analysis}}. \bibinfo{pages}{1592--1604}.
\newblock


\bibitem[Zhuo et~al\mbox{.}(2024)]%
        {zhuo2024bigcodebench}
\bibfield{author}{\bibinfo{person}{Terry~Yue Zhuo}, \bibinfo{person}{Minh~Chien
  Vu}, \bibinfo{person}{Jenny Chim}, \bibinfo{person}{Han Hu},
  \bibinfo{person}{Wenhao Yu}, \bibinfo{person}{Ratnadira Widyasari},
  \bibinfo{person}{Imam Nur~Bani Yusuf}, \bibinfo{person}{Haolan Zhan},
  \bibinfo{person}{Junda He}, \bibinfo{person}{Indraneil Paul},
  {et~al\mbox{.}}} \bibinfo{year}{2024}\natexlab{}.
\newblock \showarticletitle{Bigcodebench: Benchmarking code generation with
  diverse function calls and complex instructions}.
\newblock \bibinfo{journal}{\emph{arXiv preprint arXiv:2406.15877}}
  (\bibinfo{year}{2024}).
\newblock


\end{thebibliography}

\end{document}